\documentclass[A4paper]{article}
\usepackage{amsmath}
\usepackage{amssymb}
\usepackage{graphicx,psfrag,epsf}
\usepackage{enumerate}
\usepackage{natbib}
\usepackage{url} 
\usepackage{fullpage}
\usepackage{algorithm}
\usepackage{algorithmic}

\newcommand{\blind}{1}

\renewcommand\footnotemark{}

\newcommand{\bbeta}{\boldsymbol{\beta}}

\newcommand{\bgamma}{\boldsymbol{\gamma}}

\newcommand{\bphi}{\boldsymbol{\phi}}

\newcommand{\btheta}{\boldsymbol{\theta}}
\newcommand{\bzeta}{\boldsymbol{\zeta}}
\newcommand{\bSigma}{\boldsymbol{\Sigma}}
\newcommand{\brho}{\boldsymbol{\rho}}

\newcommand{\deltat}{\delta t}

\newcommand{\ba}{\mathbf{a}}
\newcommand{\bA}{\mathbf{A}}
\newcommand{\bd}{\mathbf{d}}
\newcommand{\bD}{\mathbf{D}}

\newcommand{\bu}{\mathbf{u}}

\newcommand{\by}{\mathbf{y}}
\newcommand{\bz}{\mathbf{z}}

\newcommand{\T}{\mathrm{T}}

\begin{document}

\def\spacingset#1{\renewcommand{\baselinestretch}%
{#1}\small\normalsize} \spacingset{1}


\if1\blind
{
  \title{\bf \Large Bayesian design of experiments for intractable likelihood models using coupled auxiliary models and multivariate emulation
  \vspace{0cm}
  }
  \author{\large Antony M. Overstall\hspace{.2cm}\\
    \large Southampton Statistical Sciences Research Institute,\\ \large University of Southampton,\\ \large Southampton, UK\\ \large (A.M.Overstall@soton.ac.uk) \\[1ex]
    \large James M. McGree\\
    \large School of Mathematical Sciences,\\ \large Queensland University of Technology,\\ \large Brisbane, Australia \\ \large (james.mcgree@qut.edu.au)}
    \date{\vspace{-1.2cm}}
  \maketitle
} \fi

\if0\blind
{
  \bigskip
  \bigskip
  \bigskip
  \begin{center}
    {\LARGE\bf Title}
\end{center}
  \medskip
} \fi

\bigskip
\begin{abstract}
A Bayesian design is given by maximising an expected utility over a design space. The utility is chosen to represent the aim of the experiment and its expectation is taken with respect to all unknowns: responses, parameters and/or models. Although straightforward in principle, there are several challenges to finding Bayesian designs in practice. Firstly, the utility and expected utility are rarely available in closed form and require approximation. Secondly, the design space can be of high-dimensionality. In the case of intractable likelihood models, these problems are compounded by the fact that the likelihood function, whose evaluation is required to approximate the expected utility, is not available in closed form. A strategy is proposed to find Bayesian designs for intractable likelihood models. It relies on the development of an automatic, auxiliary modelling approach, using multivariate Gaussian process emulators, to approximate the likelihood function. This is then combined with a copula-based approach to approximate the marginal likelihood (a quantity commonly required to evaluate many utility functions). These approximations are demonstrated on examples of stochastic process models involving experimental aims of both parameter estimation and model comparison.
\end{abstract}
\noindent%

\section{Introduction}\label{sec:intro}

Often, the dynamics underpinning a complex physical phenomenon can be modelled by a stochastic process. It is commonly the situation that the stochastic process (or model) depends on unknown parameters, time and, potentially, other controllable variables. In this paper, we consider the case where an experiment is to be performed to learn about the phenomenon by estimating the unknown parameters. That is, the physical phenomenon of interest is observed at a series of time points, after the specification of any controllable variables, and the stochastic model is \emph{fitted} to the observed responses. In particular, we focus on the optimal choice of time points and controllable variables (collectively referred to as design variables) to best learn about the unknown process. 

A feature of the stochastic models studied in this paper is that, although the dynamics behind each process can be relatively simple, the probability model linking parameters and design variables to responses is typically only defined implicitly. The development of new statistical methodology, so called \emph{likelihood-free} methodology, to analyse observed responses under these intractable likelihood models has received much attention in recent years, e.g. approximate Bayesian computation (\citealt{tavare1997}); synthetic likelihood \citep{wood2010}; variational Bayes \citep{tran_nott_kohn_2015} and auxiliary modelling (\citealt{gourieroux1993}). The task of designing the experiment, i.e. specifying the design variables, has received significantly less attention. Under the frequentist approach to statistical inference, \citet{pagendam2013} and \citet{parker2015} used numerical approximations to the Fisher information to find D-optimal designs \citep[e.g.,][Chapter 11]{atkinson_donev_tobias_2007} for stochastic epidemic and queueing models, respectively. In this paper, the Bayesian approach to statistical inference is used. Under such an approach, a utility function is specified representing the aim of the experiment. Then, a Bayesian design \citep{ChalonerVerdinelli} is found by maximising the expectation of the utility where expectation is with respect to all unknown quantities (i.e. parameters and unobserved responses) and the maximisation is over the space of all possible designs. Finding optimal designs under the Bayesian approach, even for tractable likelihood models, is a significant computational challenge; see recent reviews of the field by \cite{ryan_drovandi_mcgree_pettitt_2015} and \cite{woods_ace_2016}. Typically, neither the utility function nor its expectation are available in closed form, and the space of all possible designs can be high-dimensional. For intractable likelihood models, the problem is further exacerbated by there being no closed form expression for the likelihood. Approaches for finding Bayesian designs have been proposed that use approximate Bayesian computation (\citealt{DrovandiPettitt}, \citealt{hainy_etal_2013}, \citealt{price2016}) and auxiliary modelling (also known as indirect inference; \citealt{cmryan2016}) to approximate the likelihood. A common feature of these methodologies is that they have only been applied for examples of experiments with low-dimensional design spaces rendering them of limited practical relevance.

In this paper, we aim to overcome the shortcomings of existing approaches. The contribution is threefold. First, we apply the latest methods \citep{overstall_woods_2016} for maximising the approximate expected utility which are suitable for high-dimensional design spaces. Secondly, we develop an automatic, flexible non-parametric auxiliary modelling approach to approximate the likelihood that uses a multivariate emulator to provide an approximate link between parameters, design variables and the responses. Thirdly, we develop a novel copula-based approximation to the marginal likelihood which is a key quantity for many commonly-used utility functions and is rarely analytically tractable. The paper is organised as follows. In Section~\ref{sec:back}, we review the necessary background on Bayesian design and likelihood-free methodology. In Section~\ref{sec:meth} we describe the proposed automatic auxiliary modelling approach and the approximation to the marginal likelihood, before demonstrating the proposed methodology on illustrative yet challenging examples in Section~\ref{sec:examples}. Lastly, we describe and demonstrate how the approach can be extended to the experimental aim of model comparison in Section~\ref{sec:models}.

\section{Background}\label{sec:back}

\subsection{Intractable likelihood model} \label{sec:setup}

Suppose the experiment consists of $n$ runs. For $k=1,\dots,n$, the $k$th run involves the specification of a $w \times 1$ vector of design variables $\bd_k \in \mathcal{D}$. Let $y_k$ be the corresponding response from the phenomenon for the $k$th run. It is assumed that, independently,
\begin{equation}
y_k \sim \mathcal{F}\left(\btheta, \bd_k\right),
\label{eqn:dist}
\end{equation}
where $\mathcal{F}$ is a distribution depending on a $p \times 1$ vector of unknown parameters $\btheta \in \Theta$, with $\Theta$ the parameter space. Let $f(y|\btheta, \bd)$ denote the probability density function (pdf) or probability mass function (pmf) of the distribution $\mathcal{F}$. The likelihood is 
$$\pi(\mathbf{y}|\btheta, \bD) = \prod_{k=1}^n f(y_k|\btheta, \bd_k),$$
where $\mathbf{y} = \left(y_1,\dots, y_n \right)$ is the $n \times 1$ vector of responses, $\bD = \left(\bd_1,\dots,\bd_n\right) \in \Delta = \mathcal{D}^n$ is the vector giving the design, and $\Delta$ the $q$-dimensional design space with $q=nw$. For the models considered in this paper, $\mathcal{F}$ is only defined implicitly with the result that $f(y|\btheta, \bd)$ and $\pi(\mathbf{y}|\btheta, \bD)$ are not available in closed form. The marginal model is the distribution of $\mathbf{y}|\mathbf{D}$ having marginalised over the parameters $\btheta$. The pdf/pmf of this distribution is called the marginal likelihood (also known as evidence) and given by
\begin{equation}
\pi(\mathbf{y}|\bD) = \int_{\Theta} \pi(\mathbf{y}|\btheta,\bD) \pi(\btheta) \mathrm{d}\btheta,
\label{eqn:evidence}
\end{equation}
where $\pi(\btheta)$ is the pdf of the prior distribution for $\btheta$.

\subsection{Bayesian optimal design of experiments} \label{sec:bayesdoe}

We initially describe the concept of Bayesian optimal design of experiments for the experimental aim of parameter estimation. We consider the extension to model comparison in Section~\ref{sec:models}. Bayesian optimal design of experiments begins with the specification of a utility function denoted by $u(\btheta, \mathbf{y}, \bD)$ which represents the \emph{utility} of estimating $\btheta$ using observed responses $\mathbf{y}$ generated via design $\bD$. A Bayesian optimal design is given by maximising (over $\Delta$) the expected utility function given by
\begin{equation}
U(\bD) = \int u(\btheta, \mathbf{y}, \bD) \mathrm{dP}_{\btheta, \mathbf{y}| \bD},
\label{eqn:expu}
\end{equation}
where the expectation is with respect to the joint distribution of all unknown quantities; $\btheta$ and $\mathbf{y}$. In this paper we consider a class of utility functions which we term \emph{likelihood-based}. This is where the utility function is a functional of the likelihood, $\pi(\mathbf{y}|\btheta,\bD)$, and the marginal likelihood, $\pi(\by|\bD)$.

This class of utility function includes many commonly-employed utilities. As an example, consider the Shannon information gain (SIG; \citealt{Lindley1956}) given by
\begin{equation}
u_S(\btheta, \mathbf{y}, \bD) =  \log \pi(\mathbf{y}|\btheta,\bD) - \log \pi(\mathbf{y}|\bD).
\label{eqn:sig}
\end{equation}
The Bayesian design under the SIG utility is equivalently the design that maximises the expected (with respect to the marginal distribution of $\mathbf{y}$) Kullback Leibler divergence between the prior and posterior distributions of $\btheta$. 

Although conceptually straightforward, there are at least two hurdles to finding Bayesian designs in practice (even for tractable likelihood models). Firstly, neither the utility function $u(\btheta,\by,\bD)$ nor its expectation $U(\bD)$ are usually analytically tractable and will require approximation. Secondly, the design space $\Delta$ can be of high dimensionality, i.e. $q$ can be relatively large.

Methods proposed in the literature for approximately maximising the expected utility can be broadly classified into simulation- or smoothing-based. The simulation-based approach of \citet{Muller1999} places an artificial joint distribution on $\btheta$, $\mathbf{y}$ and $\bD$ such that the marginal pdf of $\bD$ is proportional to $U(\bD)$. Simulation methods are used to generate a sample from this joint distribution which is then used to estimate the marginal mode of $\bD$, i.e. the Bayesian design. This method has been further refined, for example, by \citet{muller_sanso_iorio_2004} and \citet{amzal_bois_parent_robert_2006}. However, difficulties in efficiently sampling over a high dimensional space mean that the typical limit of dimensionality for the design space under these methods is considered to be $q = 4$ (e.g. \citealt{ryan_drovandi_mcgree_pettitt_2015}). 

Smoothing-based approaches are based on the following Monte Carlo approximation to the expected utility
\begin{equation}
\tilde{U}(\bD) = \frac{1}{B} \sum_{i=1}^B u(\btheta_i, \mathbf{y}_i, \bD),
\label{eqn:mc}
\end{equation}
where $\left\{ \btheta_i, \mathbf{y}_i \right\}_{i=1}^B$ is a sample of size $B$ generated from the joint distribution of $\btheta$ and $\mathbf{y}$ (given $\bD$). Due to the stochastic nature of the Monte Carlo approximation, application of standard optimisation methods \citep[e.g.][]{lange_2013} is difficult. Instead, \citet{MullerParmigiani} proposed a method whereby $\tilde{U}(\bD)$ is evaluated at a series of designs and a statistical model (a smoother or emulator) is fitted that is able to predict $\tilde{U}(\bD)$ (and therefore $U(\bD)$) at any $\bD \in \Delta$. This predictor is then maximised over the design space, $\Delta$. \citet{MullerParmigiani} were able to consider design spaces with dimensionality of $q=2$. This method has been further refined by \citet{Weaver} (with maximum $q=3$) and \citet{Jones} (with maximum $q=9$). 

To aid in the applicability to design spaces of higher dimensionality, \citet{overstall_woods_2016} proposed the approximate coordinate exchange (ACE) algorithm. Here a cyclic ascent algorithm (usually referred to as coordinate exchange in the design of experiments literature; see \citealt{MeyerNachtsheim}) is used to maximise the expected utility. At each of the $q$ elements (coordinates) of the design, $\tilde{U}(\bD)$ is evaluated at a series of designs which only differ in that coordinate and a Gaussian process smoother is fitted to the resulting evaluations and used to predict $U(\bD)$ for any design. This prediction is then maximised over the one-dimensional design space of the coordinate under study. By using this methodology, \citet{overstall_woods_2016} were able to find approximately optimal designs for experiments in examples with design spaces of up to $q=192$ dimensions, i.e. nearly two orders of magnitude greater than existing methods. A brief description of the ACE algorithm is given in Section \ref{app:ace} of the Supplementary Material. Furthermore, the algorithm is implemented in the \texttt{acebayes} \citep{acebayes} \texttt{R} package. The ACE algorithm is currently the state of the art in computing Bayesian designs for realistic-sized design spaces and, for this reason, we use it in all examples. However the methodology we propose is suitable to use with any optimisation method which only requires the evaluation of the Monte Carlo approximation to the expected utility given by (\ref{eqn:mc}).

To apply any optimisation method relying on evaluation of the Monte Carlo approximation to the expected utility given by (\ref{eqn:mc}), it is a requirement to be able to evaluate the utility function $u(\btheta, \mathbf{y}, \bD)$. However, it is usually the case that the utility function itself is analytically intractable. Specifically, likelihood-based utilities depend on the marginal likelihood, e.g. the SIG utility given by (\ref{eqn:sig}), which is typically not available in closed form. The obvious approach is to use a further (inner) Monte Carlo approximation resulting in a \emph{nested Monte Carlo} approximation to the expected utility \citep{Ryan2003, Huan2013, overstall_woods_2016}. For example, to approximate the SIG utility, we generate a further sample, $\left\{ \tilde{\btheta}_j \right\}_{j=1}^C$, of size $C$ from the prior distribution of $\btheta$. The SIG utility is then approximated by 
\begin{equation}
\tilde{u}_S(\btheta, \mathbf{y}, \bD) =  \log \pi(\mathbf{y}|\btheta,\bD) - \log \tilde{\pi}(\mathbf{y}|\bD),
\label{eqn:sigmc}
\end{equation}
where the inner Monte Carlo approximation to the marginal likelihood is
\begin{equation}
\tilde{\pi}(\mathbf{y}|\bD) = \frac{1}{C} \sum_{j=1}^C \pi(\mathbf{y}|\tilde{\btheta}_j,\bD).
\label{eqn:signestedmc}
\end{equation}

\subsection{Bayesian design for intractable likelihood models} \label{sec:bayesin}

Finding Bayesian designs becomes impossible under an intractable likelihood model using the methods described in Section~\ref{sec:bayesdoe} which rely on a large number of evaluations of the likelihood $\pi(\mathbf{y}|\btheta,\bD)$ to approximate the expected utility. In the Monte Carlo approximation to the expected utility given by (\ref{eqn:mc}), the utility function is evaluated $B$ times where each evaluation of the utility needs at least $C$ evaluations of the likelihood for the inner Monte Carlo approximation to the marginal likelihood given by (\ref{eqn:signestedmc}).

We assume at this point that, although we are unable to evaluate the likelihood, it is possible to generate samples from the intractable likelihood model. All models considered in this paper are examples of Markov process models where samples can be straightforwardly generated using the Gillespie method \citep{gillespie1977}. In recent years there has been an explosion of novel methodology to evaluate the posterior distribution under an intractable likelihood depending only on the ability to generate from the model. The most popular of these methods is approximate Bayesian computation (ABC; \citealt{tavare1997}). Here, the likelihood is approximated by the ABC likelihood
$$\pi_{ABC}(\mathbf{y}|\btheta, \bD) = \int I(\delta(\tilde{\mathbf{y}},\mathbf{y}) \le \epsilon) \mathrm{dP}_{\tilde{\mathbf{y}}|\btheta, \bD},$$
where $I(A)$ is the indicator function for event $A$, $\delta(\tilde{\mathbf{y}},\mathbf{y})\ge 0$ is a discrepancy function (with $\delta(\tilde{\mathbf{y}},\mathbf{y}) = 0$ if and only if $\tilde{\mathbf{y}}=\mathbf{y}$) and $\epsilon \ge 0$ is a specified tolerance. If $\epsilon = 0$, the ABC likelihood is equal to the likelihood. The ABC likelihood is approximated via Monte Carlo. There is typically a trade-off between choosing $\epsilon$ to be sufficiently small to ensure accurate inference and large enough for computational efficiency. A similar approximation exists for the marginal likelihood given by (\ref{eqn:evidence}).

Some authors \citep[e.g.,][]{DrovandiPettitt, hainy_etal_2013, price2016} have used ABC to approximate the utility function when finding Bayesian designs. However, the ABC methodology needs the successful simultaneous specification of a discrepancy function, $\delta$, and tolerance, $\epsilon$. In inferential settings with fixed observations $\mathbf{y}$ and design, $\mathbf{D}$, it is possible to tailor these choices. However for Bayesian design, $\mathbf{y}$ is unknown meaning the choice of discrepancy function and tolerance need to be suitable for all observations under the marginal model and for any design $\mathbf{D} \in \Delta$. This means applying ABC techniques to find Bayesian designs for anything other than small $n$ is difficult \citep[e.g][]{dehideniya2017} and therefore Bayesian design for intractable likelihood models using ABC has been limited to design spaces of small dimensionality.

As mentioned in Section~\ref{sec:intro}, an alternative methodology for inference under an intractable likelihood is auxiliary modelling (also known as indirect inference). This is a well established methodology for both frequentist \citep{gourieroux1993, heggland2004} and Bayesian \citep{Drovandi2011, drovandi2015} inference.  A \emph{conditional auxiliary model} $\mathcal{F}_{X}\left(\btheta,\bd\right)$ is used to approximate the distribution, $\mathcal{F}\left(\btheta,\bd\right)$, of $y$ given in (\ref{eqn:dist}). We use the term \emph{conditional} since it is conditional on parameters $\btheta$ and to distinguish it from the marginal auxiliary model which we introduce in Section~\ref{sec:meth}. Suppose the pdf/pmf of $\mathcal{F}_{X}$ is denoted by $f_{X}(y|\btheta,\bd)$, then the auxiliary likelihood is
\begin{equation}
\pi_{X}(\mathbf{y}|\btheta,\bD) = \prod_{k=1}^n f_{X}(y_k|\btheta,\bd_k),
\label{eqn:Xlik}
\end{equation}
which is used to approximate the likelihood $\pi(\mathbf{y}|\btheta, \bD)$ in the utility function.

The conditional auxiliary model $\mathcal{F}_{X}\left(\btheta,\bd\right)$ is specified by assuming that $\mathcal{F}\left(\btheta,\bd\right) = \mathcal{H}_X\left(\bphi_f\left(\btheta,\bd\right)\right)$, where $\mathcal{H}_X(\bphi_f)$ is a probability distribution depending on $v$ \emph{auxiliary parameters}, $\bphi_f$, which are a function of the parameters $\btheta$ and design variables $\mathbf{d}$. The function $\bphi_f\left(\btheta,\bd\right)$ is estimated by generating samples from the model $\mathcal{F}\left(\btheta,\bd\right)$ under different parameters and design variables. Then $\mathcal{F}_{X}\left(\btheta,\bd\right) = \mathcal{H}_X\left(\hat{\bphi}_f\left(\btheta,\bd\right)\right)$ where $\hat{\bphi}_f$ is the estimate of $\bphi_f$.

\citet{cmryan2016} have previously used the auxiliary modelling approach to find Bayesian designs under intractable likelihood models. However they estimated $\bphi_f$ by imposing a parametric form which may lack flexibility. They also found designs using sampling-based approaches (see Section~\ref{sec:bayesdoe}) and so were restricted to design spaces of small dimensionality. In Section~\ref{sec:mgp}, we consider a more flexible non-parametric form for the function $\bphi_f$.

An additional barrier to overcome is that to consider likelihood-based utilities we also need to approximate the marginal likelihood given by (\ref{eqn:evidence}). The standard nested Monte Carlo approach (see Section~\ref{sec:bayesdoe}) would be to replace evaluation of the likelihood $\pi(\mathbf{y}|\btheta, \bD)$ by evaluation of the auxiliary likelihood $\pi_{X}(\mathbf{y}|\btheta, \bD)$ in the inner Monte Carlo approximation to the marginal likelihood given by (\ref{eqn:signestedmc}). However there exists a subtle disadvantage relating to the complexity of the conditional auxiliary model. For all pairs of $i=1,\dots,B$ and $j=1,\dots,C$, in the inner Monte Carlo approximation to the marginal likelihood we need to evaluate the auxiliary likelihood
\begin{equation}
\pi_{X}(\mathbf{y}_i|\tilde{\btheta}_j,\bD) = \exp \left( \sum_{k=1}^n \log f_{X}(y_{ik}|\tilde{\btheta}_j,\bd_k)\right),
\label{eqn:decomp}
\end{equation}
where $y_{ik}$ is the $k$th element of $\mathbf{y}_i$. In general, the exponent on the right hand side of (\ref{eqn:decomp}) can be decomposed as
\begin{equation}
\sum_{k=1}^n \log f_{X}(y_{ik}|\tilde{\btheta}_j,\bd_k) = \sum_{k=1}^n \alpha(y_{ik},\bd_k) + \sum_{k=1}^n \beta(\tilde{\btheta}_j,\bd_k) + \sum_{k=1}^n \gamma(y_{ik},\tilde{\btheta}_j,\bd_k),
\label{eqn:decomp1}
\end{equation}
for functions $\alpha$, $\beta$ and $\gamma$ whose form depend on the exact form of $f_{X}(y|\btheta,\bD)$. Therefore, to evaluate the nested Monte Carlo approximation to the expected utility, $\alpha$ and $\beta$ are evaluated $B \times n$ and $C \times n$ times each, respectively. However, $\gamma$ is evaluated $B \times C \times n$ times, which can result in a high computational burden. In cases where the nested Monte Carlo approximation has been applied previously for tractable likelihood models \citep[e.g][]{Huan2013,overstall_woods_2016}, the actual (not conditional auxiliary) model is from the exponential family of distributions and $\gamma$ can be decomposed as follows
\begin{equation}
\gamma(y, \btheta, \bd) = \gamma_y(y,\bd) \gamma_{\theta}(\btheta, \bd)
\label{eqn:decomp2}
\end{equation}
which significantly reduces the computational burden of evaluation. It transpires that $\gamma_y$ and $\gamma_{\theta}$ need only be evaluated $B \times n$ and $C \times n$ times each, respectively. However, as we demonstrate in Section~\ref{sec:examples}, the conditional auxiliary model typically needs to possess characteristics which are not found in exponential family distributions. For example, for count models, we have found that the negative binomial distribution provides a far more adequate conditional auxiliary model (see Section~\ref{sec:examples}) than the Poisson, the latter being an exponential family distribution. In these cases, the decomposition given by (\ref{eqn:decomp2}) will typically not hold. This apparently simple complication significantly increases the computational burden of evaluating the nested Monte Carlo approximation. As an alternative to nested Monte Carlo, in Section~\ref{sec:ml}, we propose an auxiliary modelling approximation to the marginal likelihood. In Section~\ref{sec:examples}, we compare designs found under nested and auxiliary Monte Carlo both in terms of accuracy and computational time.

\section{Methodology} \label{sec:meth}

\subsection{Non-parametric estimation of $\bphi_f$} \label{sec:mgp}

To estimate the function $\bphi_f$, we propose an automatic approach originating from the field of computer experiments (see, e.g., \citealt{Dean2015}, Section V). In this area, the goal is to approximate an unknown function (which is usually computationally expensive). To do this, the function is evaluated a \textquotedblleft small\textquotedblright \space number of times at a specified meta-design of arguments, and a statistical model (known as an emulator) fitted to the output. The emulator provides a prediction of the unknown function for any argument. We use the multivariate Gaussian process (MGP; \citealt{Conti2010}) model as the emulator. This is a multivariate generalisation of the Gaussian process model which is a commonly employed emulator in computer experiments. 

We begin by generating a training sample $\left\{ \mathbf{d}^{(i)} \right\}_{i=1}^M$ of size $M$ from $\mathcal{D}$. We employ the usual design used for computer experiments, i.e. a space-filling Latin hypercube design \citep[e.g.][Chapter 5]{santner_2003}. We then generate a sample $\left\{ \btheta^{(i)} \right\}_{i=1}^M$ of size $M$ from the prior distribution of $\btheta$. Finally, for $i=1,\dots, M$ we generate an independent sample $\mathbf{y}_f^{(i)} = \left(y_f^{(i1)}, \dots, y_f^{(iN)}\right)$ of size $N$ from $\mathcal{F}(\btheta^{(i)}, \mathbf{d}^{(i)})$. 

For each of these $M$ samples we compute the maximum likelihood estimate (MLE) of $\bphi_f$ under $\mathcal{H}_{X}(\bphi_f)$, i.e. let
$$\hat{\bphi}_f^{(i)} = \arg \max_{\bphi_f} \prod_{j=1}^N h_{X}(y_f^{(ij)}|\bphi_f),$$
for $i=1,\dots,M$, where $h_X(y|\bphi_f)$ is the pdf/pmf of $\mathcal{H}_X(\bphi_f)$. Typically, the MLE is not available in closed form so numerical methods are used. 

We now have $\left\{\hat{\bphi}_f^{(i)}, \mathbf{d}^{(i)}, \btheta^{(i)} \right\}_{i=1}^M$ and we learn the relationship between $\bphi_f$ and $\left(\mathbf{d},\btheta\right)$ using a MGP as follows.

Let $\mathbf{Z}_f$ be the $v \times M$ matrix where the $i$th column (for $i=1,\dots,M$) is given by $\mathbf{z}_f^{(i)} = \lambda(\hat{\bphi}_f^{(i)})$ where $\lambda$ is a monotonic and differentiable link function applied element-wise to $\hat{\bphi}_f^{(i)}$. The link function is applied so that the elements of $\mathbf{Z}_f$ are in $\mathbb{R}$, e.g. a log link if the auxiliary parameters are positive. Under the MGP, we assume that
\begin{equation}
\mathbf{Z}_f|\bbeta_f, \bSigma_f, \mathbf{A}_f \sim \mathrm{MN}\left( \bbeta_f \mathbf{1}_{M}, \mathbf{A}_f, \bSigma_f \right)
\label{eqn:mgp}
\end{equation}
where $\mathrm{MN}\left(\bbeta \mathbf{1}_{M}, \mathbf{A}_f, \bSigma_f \right)$ denotes the matrix-normal distribution with $v \times M$ mean matrix $\bbeta_f \mathbf{1}_{M}$, $v \times v$ unstructured row covariance matrix $\bSigma_f$ and $M \times M$ column correlation matrix $\mathbf{A}_f$. In (\ref{eqn:mgp}), $\mathbf{1}_{M}$ is an $1\times M$ matrix of ones, and $\bbeta_f$ is an $v \times 1$ matrix of coefficients. The $ij$th element of $\mathbf{A}_f$ is given by
\begin{equation}
A_{fij} = \kappa \left( \mathbf{x}_i, \mathbf{x}_j; \brho_f\right) + \eta_f I(i=j),
\label{eqn:A}
\end{equation}
where $\mathbf{x}_i = (\btheta^{(i)}, \mathbf{d}^{(i)})$, $\kappa(\cdot,\cdot;\brho_f)$ is a valid correlation function depending on parameters $\brho_f$, and $\eta_f>0$ is referred to as a nugget. We will assume that $\bphi_f$ is a smooth function meaning that a suitable correlation function is the following squared exponential
\begin{equation}
\kappa_{SE} \left( \mathbf{x}_i, \mathbf{x}_j; \brho_f\right) = \exp \left( - \sum_{l=1}^s \rho_{fl} \left(x_{il} - x_{jl}\right)^2 \right),
\label{eqn:sqexp}
\end{equation}
where $x_{il}$ and $\rho_{fl}$ are the $l$th elements of $\mathbf{x}_i$ and $\brho_f$, respectively. Note that $\brho_f$ is an $s \times 1$ vector where $s = p + w$. 

Suppose we wish to predict the value of $\mathbf{z}_f = \lambda(\bphi_f(\mathbf{x}))$ for any value of $\mathbf{x} = \left(\btheta, \mathbf{d}\right)$. Under (\ref{eqn:mgp}), the predictive distribution for $\mathbf{z}_f$ is given by
\begin{equation}
\mathbf{z}_f|\bbeta_f, \bSigma_f, \brho_f, \eta_f \sim \mathrm{N}\left( \bbeta_f + \left(\mathbf{Z}_f - \bbeta_f \mathbf{1}_{M}\right)\mathbf{A}_f^{-1} \mathbf{a}_f, \left(1+\eta_f - \mathbf{a}_f^{\T} \mathbf{A}_f^{-1}\mathbf{a}_f\right) \bSigma_f \right),
\label{eqn:pred}
\end{equation}
where $\mathbf{a}_f$ is an $M \times 1$ vector with $i$th element $a_{fi} = \kappa \left( \mathbf{x}_i, \mathbf{x}; \brho_f\right)$.

For simplicity, we specify that the function $\hat{\bphi}_f(\mathbf{x})$ is given by the inverse link of the predictive mean, i.e. the mean of (\ref{eqn:pred}). This depends on the parameters $\bbeta_f$, $\brho_f$ and $\eta_f$. We replace these parameters by their MLEs where the likelihood is given by (\ref{eqn:mgp}). Thus 
\begin{equation}
\hat{\bphi}_f(\mathbf{x}) = \lambda^{-1}\left(\hat{\bbeta}_f + \left(\mathbf{Z}_f - \hat{\bbeta}_f\mathbf{1}_{M}\right)\hat{\mathbf{A}}_f^{-1} \hat{\mathbf{a}}_f\right),
\label{eqn:predmeanMGP}
\end{equation}
where $\hat{\mathbf{a}}_f$ and $\hat{\mathbf{A}}_f$ are $\mathbf{a}_f$ and $\mathbf{A}_f$, respectively, with $\brho_f$ and $\eta_f$ replaced by their MLEs, $\hat{\brho}_f$ and $\hat{\eta}_f$, respectively. 

Finding the conditional auxiliary model using the above approach does carry high computational burden. However it can be entirely completed \emph{off-line}, i.e. prior to starting any algorithm for maximising the approximate expected utility. Although we have prescribed an automatic approach to estimating the function $\bphi_f$, we still need to specify the distribution $\mathcal{H}_X$.  In Section~\ref{sec:assess}, we propose methods for assessing the adequacy of auxiliary models. We advocate an iterative approach whereby an auxiliary model is fitted, assessed for adequacy and, in light of this assessment, potentially updated, i.e. the same approach one would use for statistical modelling of physical data. 

\subsection{Approximating the marginal likelihood} \label{sec:ml}

For $k=1,\dots,n$, let $\mathcal{G}(\mathbf{d}_k)$ be the marginal distribution of $y_k$ having marginalised over the parameters $\btheta$, i.e. the pdf/pmf of $\mathcal{G}(\mathbf{d}_k)$ is given by
$$g(y_k|\mathbf{d}_k) = \int_{\Theta} f(y_k|\btheta, \mathbf{d}_k) \pi(\btheta) \mathrm{d}\btheta.$$
If the elements of $\mathbf{y}$ are continuous then, by Sklar's theorem \citep[e.g.][Section 2.3]{nelson1998}, the marginal likelihood is uniquely given by
\begin{equation}
\pi(\mathbf{y}|\bD) = c\left(G(y_1|\bd_1), \dots, G(y_n|\bd_n)|\bD\right) \times \prod_{k=1}^n g(y_k|\bd_k),
\label{eqn:copula}
\end{equation}
where $c$ is the pdf of the copula $\mathcal{C}$ of the marginal model, $\mathbf{y}|\mathbf{D}$, and $G(y_k|\mathbf{d}_k)$ is the cumulative distribution function (cdf) of $\mathcal{G}(\mathbf{d}_k)$, for $k=1,\dots,n$. Suppose that $\mathbf{u} = \left(G(y_1|\bd_1), \dots, G(y_n|\bd_n)\right)$, then the marginal distribution of each element of $\mathbf{u}$ is $\mathrm{U}[0,1]$ and the copula is the joint distribution of $\mathbf{u}$. Essentially a continuous multivariate probability distribution can be decomposed into the marginal distributions of each element and the copula which controls the dependency structure. 

Suppose we find a suitable auxiliary model (termed the \emph{marginal auxiliary model}), denoted by $\mathcal{G}_{X}(\bd)$, for $\mathcal{G}(\bd)$, with pdf $g_{X}(y|\bd)$ and cdf $G_{X}(y|\bd)$, and an \emph{auxiliary copula}, $\mathcal{C}_{X}$, with pdf $c_{X}(\mathbf{u}|\bD)$, then the marginal likelihood can be approximated using
\begin{equation}
\pi_{X}(\mathbf{y}|\bD) = c_{X}\left(G_{X}(y_1|\bd_1), \dots, G_{X}(y_n|\bd_n)|\bD\right) \times \prod_{k=1}^n g_{X}(y_k|\bd_k).
\label{eqn:iicopula}
\end{equation}
The decomposition given by (\ref{eqn:copula}) is only unique for continuous $\by$. However, this does not preclude its use for discrete $\by$ \citep[see, e.g.,][]{pana2012}.

The marginal auxiliary model is constructed in an analogous way to the construction of the conditional auxiliary model, i.e. we assume that $\mathcal{G}(\mathbf{d}) = \mathcal{H}_X(\bphi_g(\mathbf{d}))$ and we set $\mathcal{G}_X(\mathbf{d}) = \mathcal{H}_X(\hat{\bphi}_g(\mathbf{d}))$ where $\hat{\bphi}_g$ is the estimate of $\bphi_g$. The estimate $\hat{\bphi}_g$ is found using a MGP in a similar way to how we estimate $\bphi_f$ in Section~\ref{sec:mgp}. Full details are given in Section~\ref{app:margaux} of the Supplementary Material. Note that similar to the construction of $\hat{\bphi}_f$, we complete this off-line, prior to starting any algorithm for maximising the approximate expected utility.  

\subsection{Constructing the auxiliary copula} \label{sec:copula}

We now consider specifying the auxiliary copula. Similar to forming the conditional and marginal auxiliary models, we choose a family for the copula depending on an $r \times 1$ vector of \emph{copula parameters}, $\bzeta$. In contrast to the conditional and marginal auxiliary models, the choice of copula family will be less intuitive. However, for all examples in this paper, the t-copula \citep[e.g.][]{demartamcneil2005} sufficed to produce an adequate coupled auxiliary model. We compared to the simpler normal copula \citep[e.g.][]{joe1997} and found negligible difference in terms of approximate expected utility of the designs found. However we favour the more complex t-copula for its flexibility in accounting for extreme values. Unlike the conditional and marginal auxiliary models, we propose that the specification of the auxiliary copula be made \emph{on-line}, i.e. during evaluation of $\tilde{U}(\bD)$ within the chosen algorithm for the maximisation of the approximate expected utility. The reasoning for this difference is as follows. In the case of the conditional and marginal auxiliary models, the dimensionality of the arguments of the functions $\bphi_f$ and $\bphi_g$ are $p+w$ and $w$, respectively, i.e. relatively small. However, if we were to allow $\bzeta$ to be a function of $\mathbf{D}$, the dimensionality of this argument is $q=nw$, i.e. relatively large, for which it may not be possible to estimate $\bzeta$ reliably for all $\mathbf{D} \in \Delta$. At each evaluation of $\tilde{U}(\bD)$, since $\bD$ is fixed, $\bzeta$ is independent of $\bD$ and its value estimated using a copula training sample generated from the model. Therefore, we write the copula as $\mathcal{C}_{X}(\bzeta)$ with pdf $c_{X}(\mathbf{u}|\bzeta)$, i.e. independent of $\mathbf{D}$. The pdf for the t-copula with $\delta$ degrees of freedom at $\mathbf{u} = \left(u_1,\dots,u_n\right)$ is 
\begin{equation}
c_{X}(\mathbf{u}|\bzeta) = |\mathbf{R}(\bgamma)|^{-\frac{1}{2}} \left[ \frac{ \delta + \mathbf{v}^T\mathbf{v}}{\delta + \mathbf{v}^T \mathbf{R}(\bgamma)^{-1} \mathbf{v}} \right]^{\frac{\delta + n}{2}}.
\label{eqn:tcopula}
\end{equation}
In (\ref{eqn:tcopula}), $\mathbf{v}$ is an $n \times 1$ vector with $k$th element $v_k = T_{\delta}^{-1}(u_k)$, $T_{\delta}$ is the distribution function of the standard univariate t-distribution with $\delta$ degrees of freedom, and $\mathbf{R}(\bgamma)$ is an $n \times n$ correlation matrix with $\frac{1}{2}n(n-1)$ unique elements given by the elements of $\bgamma$. The $r = \frac{1}{2}n(n-1) + 1$ copula parameters, given by $\bzeta = \left( \bgamma, \delta \right)$, are estimated via a two-stage process where $\bgamma$ is estimated via method of moments and $\delta$ by maximum likelihood \citep[e.g.][]{demartamcneil2005}. 

\subsection{Assessing adequacy of auxiliary models} \label{sec:assess}

Before applying the auxiliary models described in the previous section to approximate the expected utility, their adequacy should be assessed for plausibility, i.e. do they provide a reasonable approximation to the assumed model. The approach proposed is based on posterior predictive assessments (see, for example, \citealt{gelman_etal_2014}, Chapter 6). Here $M_0$ test samples are generated from the assumed and auxiliary models, and sample statistics from each compared. We propose to separately assess a) the conditional and marginal auxiliary models and; b) the coupled auxiliary model.

\subsubsection{Assessing the conditional and marginal auxiliary models} \label{sec:condmarglik}

We generate $M_0$ samples of size $N$ from the assumed, $\left\{ \bar{\by}^{(i)}_f \right\}_{i=1}^{M_0}$, and auxiliary conditional, $\left\{ \bar{\by}^{(i)}_{fX} \right\}_{i=1}^{M_0}$, models using the following two steps.

\begin{enumerate}
\item
Generate test samples, $\left\{ \bar{\btheta}^{(i)} \right\}_{i=1}^{M_0}$ and $\left\{ \bar{\bd}^{(i)} \right\}_{i=1}^{M_0}$, of size $M_0$ from the prior distribution of $\btheta$ and uniformly over $\mathcal{D}$, respectively.
\item
For $i=1,\dots,M_0$ and $j=1,\dots,N$ generate
$$\bar{y}_f^{(ij)} \sim \mathcal{F}\left( \bar{\btheta}^{(i)}, \bar{\bd}^{(i)}\right) \qquad \bar{y}_{fX}^{(ij)} \sim  \mathcal{H}_{X}\left( \hat{\bphi}_f\left(\bar{\btheta}^{(i)},\bar{\bd}^{(i)}\right)\right).$$
Let
$$\bar{\by}_f^{(i)} = \left(\bar{y}_f^{(i1)},\dots,\bar{y}_f^{(iN)}\right) \qquad \bar{\by}_{fX}^{(i)} =  \left(\bar{y}_{fX}^{(i1)},\dots,\bar{y}_{fX}^{(iN)}\right).$$
\end{enumerate}

We propose two diagnostics to compare these samples. First, plot sample statistics of the $\bar{\by}_f^{(i)}$'s against $\bar{\by}_{fX}^{(i)}$'s, where suggested sample statistics are mean, variance, median, etc. If the conditional auxiliary model is adequate then the points should approximately lie on a straight line through the origin with unit slope. Second, a single number summary of conditional auxiliary model adequacy is given by the Bayesian posterior predictive p-value \citep[][page 146]{gelman_etal_2014} given by

$$\mbox{p-value}_f = \frac{1}{M_0} \sum_{i=1}^{M_0} I \left(\sum_{j=1}^N \log h_{X} \left( \bar{y}_f^{(ij)} | \hat{\bphi}_f(\bar{\btheta}^{(i)},\bar{\bd}^{(i)}) \right) < \sum_{j=1}^N \log h_{X} \left( \bar{y}_{fX}^{(ij)} | \hat{\bphi}_f(\bar{\btheta}^{(i)},\bar{\bd}^{(i)}) \right)\right).$$
A posterior predictive p-value close to zero or one indicate that the auxiliary model is inadequate. Similar diagnostics can be obtained for the marginal auxiliary model.

\subsubsection{Assessing the coupled auxiliary model}

To assess the adequacy of the coupled auxiliary model, we generate $M_0$ samples of size $n$ from both the marginal model, $\left\{\breve{\mathbf{y}}^{(i)}\right\}_{i=1}^{M_0}$, and the coupled auxiliary model, $\left\{\breve{\mathbf{y}}_X^{(i)}\right\}_{i=1}^{M_0}$. The steps required to generate these samples is given in Section~\ref{app:couple} of the Supplementary Material. To compare these samples, we use a posterior predictive p-value given by 		
$$\mbox{p-value} = \frac{1}{M_0} \sum_{i=1}^{M_0} I\left( \log \pi_{X}\left(\breve{\mathbf{y}}^{(i)} | \breve{\bD}^{(i)}\right) > \log \pi_{X}\left(\breve{\mathbf{y}}_{X}^{(i)} | \breve{\bD}^{(i)}\right) \right).$$
Similar to assessing the conditional and marginal auxiliary models, a posterior predictive p-value close to zero or one suggests an inadequate coupled auxiliary model.

\subsection{The auxiliary Monte Carlo approximation to the expected likelihood-based utility} \label{sec:exputility}	

We now summarise the steps required to approximate the expected utility given a design $\bD = \left(\bd_1,\dots,\bd_n\right)$, a Monte Carlo sample size $B$ and a copula training sample size $L$. Note that these steps rely on the \emph{off-line} construction of both the conditional auxiliary model $\mathcal{F}_X(\btheta, \mathbf{d})$ (with pdf/pmf $f_X(y|\btheta, \mathbf{d})$) and the marginal auxiliary model $\mathcal{G}_X(\mathbf{d})$ (with pdf/pmf $g_X(y|\mathbf{d})$ and distribution function $G_X(y|\mathbf{d})$).

\begin{enumerate}
\item
Generate sample, $\left\{ \btheta_i \right\}_{i=1}^B$ from the prior distribution of $\btheta$. For $i=1,\dots,B$ and $k=1,\dots,n$, generate
$$y_{ik} \sim \mathcal{F}\left(\btheta_i, \bd_k \right),$$
and let $\mathbf{y}_i = \left(y_{i1},\dots,y_{in}\right)$. Now $\left\{\mathbf{y}_i, \btheta_i \right\}_{i=1}^B$ is the Monte Carlo sample from the joint distribution of $\by$ and $\btheta$ given $\bD$.
\item
Generate sample, $\left\{ \bar{\btheta}_i \right\}_{i=1}^L$ from the prior distribution of $\btheta$. For $l=1,\dots,L$ and $k=1,\dots,n$, generate
$$\bar{y}_{lk} \sim \mathcal{F}\left(\bar{\btheta}_l, \bd_k \right),$$
and let $\bar{\mathbf{y}}_l = \left(\bar{y}_{l1},\dots,\bar{y}_{ln}\right)$. Now $\left\{ \bar{\by} \right\}_{l=1}^L$ is the copula training sample from the marginal distribution of $\mathbf{y}$ given $\bD$.
\item
Calculate the maximum likelihood estimates, $\hat{\bzeta}$, of $\bzeta$ where
$$\hat{\bzeta} = \arg \max_{\bzeta} \prod_{l=1}^L c_{X} \left( G_{X}(\bar{y}_{l1}|\bd_1), \dots, G_{X}(\bar{y}_{ln}|\bd_n) | \bzeta \right).$$
This maximisation will need to be computed numerically since closed form maximum likelihood estimates typically do not exist for copula parameters. 
\item
For $i=1,\dots,B$, calculate the following approximations to the likelihood and marginal likelihood
\begin{eqnarray*}
\pi_{X}(\mathbf{y}_i|\btheta_i,\bD) & = & \prod_{k=1}^n f_{X}(y_{ik}|\btheta_i,\bd_k),\\
\pi_{X}(\mathbf{y}_i|\bD) & = & c_{X}\left( G_{X}(y_{i1}|\bd_1), \dots, G_{X}(y_{in}|\bd_n) | \hat{\bzeta}\right) \prod_{k=1}^n g_{X}(y_{ik}|\bd_k).
\end{eqnarray*}
\item
For $i=1,\dots,B$, approximate the likelihood-based utility, $u(\btheta_i,\mathbf{y}_i,\bD)$, by $u_X(\btheta_i,\mathbf{y}_i,\bD)$, wherein the likelihood and marginal likelihood are replaced by $\pi_X(\mathbf{y}_i|\btheta_i, \bD)$ and $\pi_X(\mathbf{y}_i|\bD)$, respectively. The resulting approximation to the expected utility, given by
$$\tilde{U}(\mathbf{D}) = \frac{1}{B}\sum_{i=1}^B u_X(\mathbf{y}_i,\btheta_i,\bD),$$
is termed the \emph{auxiliary Monte Carlo} approximation to the expected utility.
\end{enumerate}

\section{Examples} \label{sec:examples}

We apply the proposed methodology on a series of examples. To demonstrate the methodology and assess its efficacy, in Section~\ref{sec:comp}, we consider an illustrative example of the compartmental non-linear model where the likelihood is available in closed form. We then apply the methodology to an aphid population growth model (Section~\ref{sec:aphids}) and a parasite model (Section~\ref{sec:cats}), both of which have been used in the literature to demonstrate Bayesian design under intractable likelihood. 

First we describe some implementation details common to all examples. For the training samples, we set $M=500$ (number of marginal and conditional auxiliary model training samples), $N=10000$ (size of training sample size) and $L=500$ (number of copula training samples). These were found to be sufficient in all examples to provide adequate auxiliary models. To assess adequacy, we used $M_0=100$ test samples. To assess the coupled auxiliary model, we compute the posterior predictive p-value for all values of $n$ considered for each example.

\subsection{Compartmental model} \label{sec:comp}

In this section we apply the proposed methods to find a Bayesian design under the SIG utility for a compartmental model. For this model, the likelihood is available in closed form so the aim of this example is to assess the efficacy of the approach. Compartmental models simulate how materials flow through an organism. The design problem is to specify the $n$ sampling times $\bD = \left(t_1,\dots,t_n\right)$ (in hours) at which to measure the concentration of a drug in an individual, following the administration of the drug at time $t=0$. The concentration at time $t_k$ is denoted by $y_k$ where it is assumed that
$$y_k \sim \mathrm{N}\left( \mu(\btheta;t_k), \nu(\btheta;t_k)\right),$$
with $\btheta = \left(\theta_1,\theta_2,\theta_3\right)$ being the unknown parameters,
$$\mu(\btheta;t_k) =  \frac{400 \theta_2}{\theta_3\left(\theta_2 - \theta_1\right)} \left( \exp \left( - \theta_1 t_k \right) - \exp \left( - \theta_2 t_k \right) \right), \qquad \nu(\btheta;t_k) = 0.1 + 0.01\mu(\btheta;t_k)^2,$$
and $n=15$. Following \cite{ryan_drovandi_thompson_pettitt_2014}, independent prior distributions are assumed for the elements of $\btheta$, where, on the log scale, the common variance is 0.05 and the expectations are $\log(0.1)$, $\log(1)$ and $\log(20)$, respectively. Additionally, a constraint is imposed on the design whereby sampling times must be at least 15 minutes apart. \cite{overstall_woods_2016} describe how such constraints can be easily incorporated into the ACE algorithm.

For the distribution, $\mathcal{H}_{X}$, we use the normal distribution dependent on $v=2$ auxiliary parameters, $\bphi  = \left(\phi_1,\phi_2\right)$ controlling the mean and variance, respectively. The variance parameter, $\phi_2$, is positive so the $\lambda$ link function is chosen to be the log function for this element. After fitting the auxiliary models, the posterior predictive p-values associated with the conditional and marginal auxiliary models are $\mbox{p-value}_f = 0.62$ and $\mbox{p-value}_g= 0.42$. The posterior predictive p-value associated with the coupled auxiliary model is 0.47. Figure~\ref{fig:comp_check} in the Supplementary Material shows plots of sample statistics (mean and variance) of the $\by^{(i)}_f$'s (the $\by^{(i)}_g$'s) against the $\by^{(i)}_{fX}$'s (the $\by^{(i)}_{gX}$'s). These plots and the posterior predictive p-values show that the auxiliary models appear adequate.

\begin{table}
\centering
\caption{Mean (standard error) nested Monte Carlo approximation (under the exact likelihood) to the expected SIG utility for the compartmental model under the designs found under the four different approaches.} \label{tab:comp_perf}
\begin{tabular}{ll}
Approach & Mean (standard error) \\ \hline 
Nested Monte Carlo (exact likelihood)& 4.51 (0.003) \\
Auxiliary Monte Carlo & 4.28 (0.003) \\
Auxiliary Monte Carlo (exact likelihood) & 4.48 (0.003) \\
Equally-spaced design & 3.70 (0.003) \\  \hline
\end{tabular}
\end{table} 

We find Bayesian designs under the SIG utility using ACE under two different approaches. In the first, we use the auxiliary Monte Carlo approximation to the expected utility, as described in Section~\ref{sec:meth}, where we approximate both the likelihood and marginal likelihood using the auxiliary models. In the second approach, we only approximate the marginal likelihood, using the coupled auxiliary model, since the likelihood for the compartmental model is available in closed form. We compare the resulting two designs to a) the SIG design found by using nested Monte Carlo (under the exact likelihood) to approximate the expected utility; and b) the design given by equally-spaced sampling times. The former was found by \cite{overstall_woods_2016} using the ACE algorithm. Table~\ref{tab:comp_perf} shows the mean and standard error of twenty nested Monte Carlo approximations (under the exact likelihood) to the expected utility under each of the four designs. The design found under the methodology proposed in this paper for intractable likelihood models (i.e. second row of Table~\ref{tab:comp_perf}) performs reasonably. Obviously being able to evaluate the exact likelihood (third row of Table~\ref{tab:comp_perf}) improves this design to a point that it has performance close to the design found under nested Monte Carlo with the exact likelihood. We conclude that the methodology is competitive.

\begin{figure}
\begin{center}
\caption{Graphics for assessing the adequacy of the auxiliary model for the aphid model. In the first row, (a) shows a plot of the sample mean of the $\by_{fX}^{(i)}$'s against the sample mean of the $\by_{f}^{(i)}$'s for the Poisson conditional auxiliary model. Corresponding plots for the negative binomial conditional (b) and marginal (c) auxiliary models. The second row shows the corresponding plots for the sample variance.} \label{fig:aphids_check}
\includegraphics[scale=0.75]{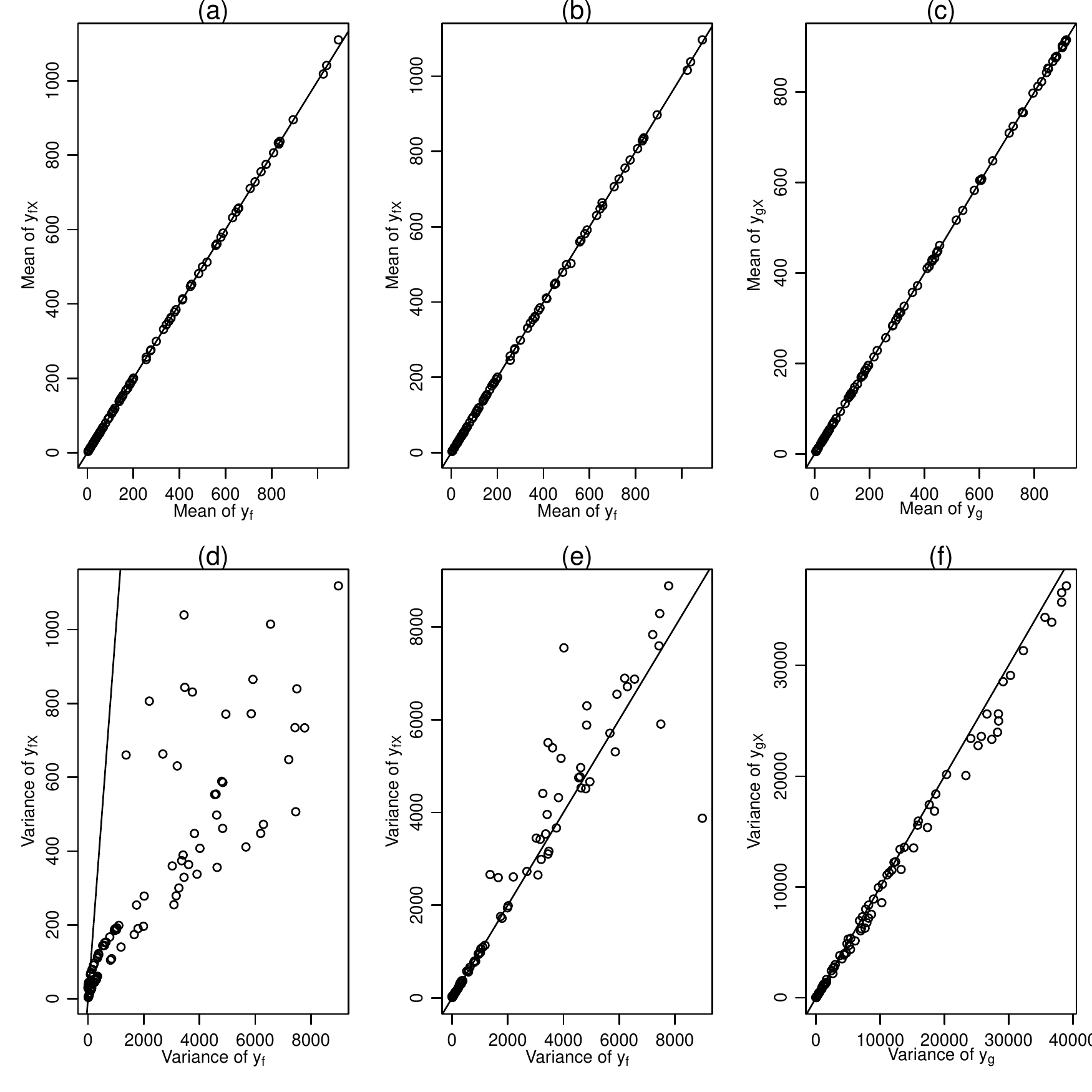}
\end{center}
\end{figure}

\subsection{Aphid population growth model} \label{sec:aphids}

 \begin{table}
\centering
\caption{Average computing time (in hours) for designs found under auxiliary and nested Monte Carlo for the SIG utility and the aphid model.} \label{tab:aphids_times}
\begin{tabular}{lcccccccccc}
Number of runs, $n$ & 5 & 10 & 15 & 20 & 25 & 30 & 35 & 40 & 45 & 50 \\ \hline
Auxiliary Monte Carlo & 0.5 & 1.7 & 3.7 & 6.2 & 9.6 & 13.6 & 18.2 & 23.8 & 29.7 & 36.1 \\
Nested Monte Carlo & 8.5 & - & - & - & - & - & - & - & - & -\\ \hline
\end{tabular}
\end{table} 

Now consider an experiment to learn about aphid infestation in cotton plants. In the experiment, the number of aphids, $y_k$, in a plot of cotton plants is recorded at sampling time $t_k$ (in days), for $k=1,\dots,n$. Let $N(t)$ and $C(t)$ denote the number of current and cumulative aphid population sizes at time $t$. \citet{matis2007} proposed a Markov model for the aphid population where the dynamics are given by the following equations:
\begin{eqnarray*}
P(N(t + \deltat) = N(t) + 1, C(t + \deltat) = C(t) + 1) &=& \theta_1 N(t) \deltat + o(\deltat),\\
P(N(t + \deltat) = N(t) - 1, C(t + \deltat) = C(t) ) &=& \theta_2 N(t)C(t) \deltat + o(\deltat),
\end{eqnarray*}
where $\btheta = \left(\theta_1,\theta_2\right)$ are the unknown parameters. Therefore the population of aphids experiences a birth rate of $\theta_1 N(t)$ and a death rate of $\theta_2 N(t)C(t)$. Note that $y_k = N(t_k)$ and the design is $\bD = \left(t_1,\dots,t_n\right)$. Finding designs for this experiment was considered by \citet{gillespieboys2018} under a non-likelihood-based utility given by the determinant of the posterior precision matrix for $\btheta$. They used a moment closure approach to approximate the stochastic model \citep{gillespiegolightly2010} but only considered experiments with a low-dimensional design space. Following \citet{gillespieboys2018}, we assume \emph{a-priori} that 
$$\btheta \sim \mathrm{N}\left( \left( \begin{array}{c}
2.46 \times 10^{-1}\\
1.34 \times 10^{-4} \end{array}\right), \left( \begin{array}{cc} 6.24 \times 10^{-5} & 5.80 \times 10^{-8} \\
5.80 \times 10^{-8} & 4.00 \times 10^{-10} \end{array}\right)\right),$$
let $N(0) = C(0) = 28$ and $\Delta = [0,49]$ days. Finally, we consider a range of different experiment sizes, i.e. $n = 5, 10, \dots, 50$.

Since the response is a count, the natural choice for $\mathcal{H}_X$ is a Poisson distribution. This has $v=1$ auxiliary parameter giving both the mean and variance of the auxiliary model. Since this parameter is positive we employ the log link. We fit the conditional auxiliary model and assess its adequacy. The first column of Figure~\ref{fig:aphids_check} shows plots of the sample mean (a) and sample variance (d) of the $\by_{fX}^{(i)}$'s against the $\by_f^{(i)}$'s. Whereas there appears to be good agreement between the means of the two models, the conditional auxiliary model appears to be severely underestimating the variance. Instead we use a negative binomial distribution which is a common alternative to the Poisson distribution in the presence of over-dispersion. The negative binomial distribution has $v=2$ positive auxiliary parameters so again the log link is used. The second column of Figure~\ref{fig:aphids_check} shows plots of the sample mean (b) and sample variance (e) of the $\by_{fX}^{(i)}$'s against the $\by_f^{(i)}$'s. The third column shows the corresponding plots for the $\by_{gX}^{(i)}$'s against the $\by_g^{(i)}$'s. Clearly there is now agreement between both the sample means and the sample variances. The posterior predictive p-values for the conditional and marginal auxiliary models are $\mbox{p-value}_f = 0.43$ and $\mbox{p-value}_g = 0.37$, respectively. For the range of different values of $n$, the posterior predictive p-values for the coupled auxiliary model correspondingly ranged from 0.24 to 0.47. We conclude that the auxiliary models are adequate.

We consider finding Bayesian designs under the SIG utility and an alternative likelihood-based utility that we term \emph{likelihood ratio} (LR), given by 
$$u_{LR}(\btheta, \mathbf{y}, \mathbf{d}) = 1 - \pi(\mathbf{y}|\mathbf{d})^{\frac{1}{2}}\pi(\mathbf{y}|\btheta,\mathbf{d})^{-\frac{1}{2}}.$$
The design that maximises the expected LR utility, equivalently maximises the expected Hellinger distance between the prior and posterior distributions. 


We find designs using ACE under auxiliary Monte Carlo for each value of $n$ and each utility function. We compare these designs against the design formed from $n$ equally spaced sampling times. Finally, designs are also found under each utility for $n=5$ using nested Monte Carlo (using the auxiliary likelihood). Only $n=5$ was considered due to the computational expense of finding designs using nested Monte Carlo for $n>5$. 

\begin{figure}
\begin{center}
\caption{Plots summarising results from the aphid model. Plots (a) and (b) show the mean nested Monte Carlo approximation (under auxiliary likelihood) to the expected SIG and LR utilities, respectively, against $n$ for designs found under the three different approaches. Plot (c) shows 200 samples of the aphid population size generated from the aphid model plotted against time. Plots (d) and (e) show the SIG and LR designs, respectively, found under auxiliary Monte Carlo for each value of $n$.}
\label{fig:aphids_perf}
\includegraphics[scale=0.75]{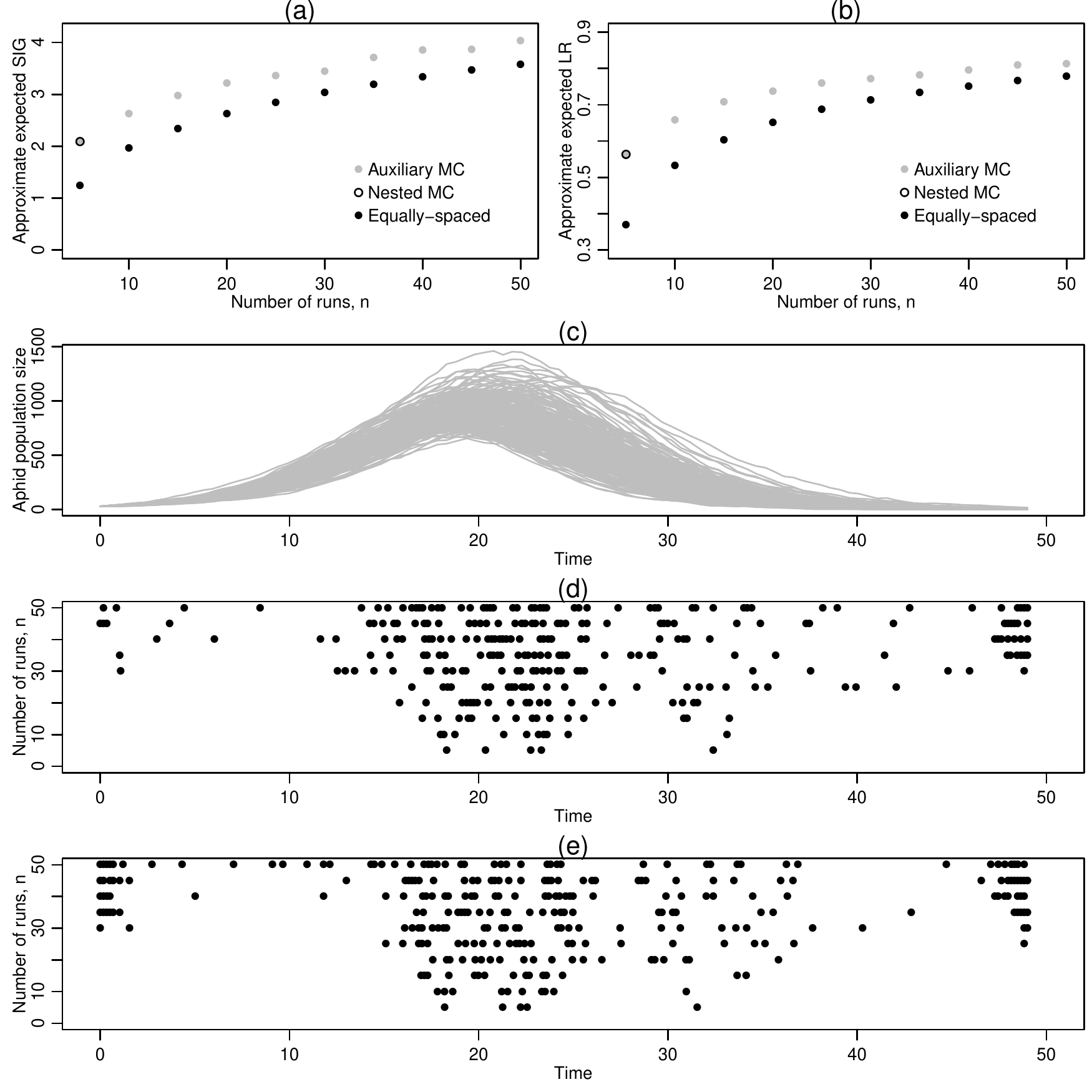}
\end{center}
\end{figure}

The first row of Figure~\ref{fig:aphids_perf} shows the mean of twenty nested Monte Carlo approximations (under auxiliary likelihood) to the expected (a) SIG and (b) LR utilities against $n$ for the three different types of design. As expected, in both cases, as $n$ increases the expected gain in utility increases. The SIG and LR designs found under auxiliary Monte Carlo are superior to the equally-spaced designs, and, for $n=5$, there is negligible difference between the designs found under nested and auxiliary Monte Carlo. 

Table~\ref{tab:aphids_times} shows the average computing times required to find the designs under the SIG utility for auxiliary and nested Monte Carlo for each value of $n$. The times for the LR utility are similar. For $n=5$, it can be seen that finding the nested Monte Carlo design requires over 15 times as much computing time relative to the auxiliary Monte Carlo design. The additional computational expense of the nested Monte Carlo approximation can be explained by considering the decomposition given in (\ref{eqn:decomp1}). Under the negative binomial auxiliary model, $\alpha(y,\mathbf{d}) = - y!$, $\beta(\btheta, \mathbf{d}) = \hat{\phi}_{f2} \log \hat{\phi}_{f2} - \log \Gamma \left( \log \hat{\phi}_{f2}\right)$ and 
$$\gamma(y,\btheta,\mathbf{d}) = \log \Gamma(\hat{\phi}_{f2} + y) + y \hat{\phi}_{f1} - (y + \hat{\phi}_{f2}) \log (\hat{\phi}_{f1} + \hat{\phi}_{f2}),$$
where $\left( \hat{\phi}_{f1}, \hat{\phi}_{f2}\right) = \hat{\bphi}_f(\btheta,\mathbf{d})$ are the $v=2$ estimated auxiliary parameters under the conditional auxiliary model. The function $\gamma(y,\btheta,\mathbf{d})$ cannot be written in the form of (\ref{eqn:decomp2}) due to the Gamma function and the same is true even after applying the Stirling approximation to the Gamma function \citep[e.g.][page 257]{abra_ste}. 

Figure~\ref{fig:aphids_perf} (c) shows a plot of 200 samples of the aphid population size generated from the aphid model plotted against time. Figures~\ref{fig:aphids_perf} (d) and~\ref{fig:aphids_perf} (e)  shows the SIG and LR designs, respectively, found under auxiliary Monte Carlo for each value of $n$. It can be seen in both cases, that for small values of $n$, the designs have sampling times concentrated in the middle of the sampling window corresponding to the \textquotedblleft peak" in the aphid population. This qualitatively agrees with the designs found by \citet{gillespieboys2018} under their non-likelihood-based utility function. However, as $n$ increases, we find the designs also include sampling times at the extremes of the sampling window.

\subsection{Parasite model} \label{sec:cats}

We now consider a parasite model example modified from \cite{DrovandiPettitt} and \cite{cmryan2016}. In the experiment, the $k$th host cat is injected with $d_{k1} \in [100,200]$ \emph{Brugia pahangi} larvae at time $t=0$, for $k=1,\dots,n$. After time $d_{k2} \in (30,300)$ (in days), the $k$th host cat is sacrificed and the number of mature parasites, $y_k$, are counted at autopsy. \cite{riley2003} proposed a Markov process to simulate the population of parasites within the host cat. At time $t$, let $J(t)$ and $M(t)$ denote the number of juvenile and mature parasites, respectively. Furthermore, let $I(t)$ be a discrete representation of the host immunity. The dynamics of the model are as follows
\begin{eqnarray*}
& & \mathrm{P}\left(J(t+\deltat) = J(t) - 1, M(t+\deltat) = M(t) + 1, I(t+\deltat) = I(t)\right) = \theta_1 J(t) \deltat + o(\deltat),\\
& & \mathrm{P}\left(J(t+\deltat) = J(t) - 1, M(t+\deltat) = M(t) , I(t+\deltat) = I(t)\right) \\
& & \qquad \qquad \qquad \qquad \qquad \qquad \qquad \qquad \qquad \qquad \quad = \left(\theta_4 + \theta_5 I(t)\right) J(t)\deltat +o(\deltat),\\
& & \mathrm{P}\left(J(t+\deltat) = J(t) , M(t+\deltat) = M(t)-1 , I(t+\deltat) = I(t)\right) = \theta_2M(t)\deltat  +o(\deltat),\\
& & \mathrm{P}\left(J(t+\deltat) = J(t) , M(t+\deltat) = M(t), I(t+\deltat) = I(t)+1\right) = \theta_3J(t)\deltat  +o(\deltat),\\
& & \mathrm{P}\left(J(t+\deltat) = J(t) , M(t+\deltat) = M(t), I(t+\deltat) = I(t)-1\right) = \theta_6I(t)\deltat  +o(\deltat),
\end{eqnarray*}
where $\btheta = \left(\theta_1,\dots,\theta_6\right)$ are unknown parameters. Juvenile and mature parasites die with rates $\left(\theta_4 + \theta_5 I(t)\right) J(t)$ and $\theta_2M(t)$, respectively. Juvenile parasites mature with rate $\theta_1J(t)$. The discrete measure of cat immunity increases or decreases by one unit with rates $\theta_3J(t)$ or $\theta_6I(t)$, respectively. 

Note that for the $k$th run, $J(0) = d_{k1}$, $M(0)=0$, $I(0)=0$ and $M(d_{k2})=y_k$. Both \cite{DrovandiPettitt} and \cite{cmryan2016} fixed all parameter values except $\theta_3$ and $\theta_4$ and considered a design space with a maximum dimensionality of four. This was either by setting $n=4$ and fixing $d_{k1}=200$ or setting $n=2$. We consider all elements of $\btheta$ to be unknown and consider number of runs $n=2,4,6,8,10,20,30,40$, thus considering a design space with a maximum dimensionality of 80. The prior distributions for $\btheta$ follow from the analysis of responses from a previous experiment with $n=212$ \citep{denham1977}. Following \cite{DrovandiPettitt}, the prior distribution for $\theta_3$ and $\theta_4$ is given by 
$$\left(\begin{array}{c}
\sqrt{\theta_3}\\
\sqrt{\theta_4}\end{array}\right) \sim \mathrm{N}\left(\left( \begin{array}{c} 
0.0361\\
0.0854 \end{array}\right), \left( \begin{array}{cc}
2.03 \times 10^{-5} & - 1.07 \times 10^{-4} \\
- 1.07 \times 10^{-4} & 1.17 \times 10^{-3} \end{array} \right)\right).$$
The remaining parameters are given Gamma prior distributions with 
\begin{equation}
\begin{array}{lcllcl}
\mathrm{E}\left(\theta_1\right) & = & 0.04 & \mathrm{var}\left(\theta_1\right) & = & 4.00 \times 10^{-4}\\
\mathrm{E}\left(\theta_2\right) & = & 0.00147 & \mathrm{var}\left(\theta_2\right) & = & 2.56 \times 10^{-7}\\
\mathrm{E}\left(\theta_5\right) & = & 1.10 & \mathrm{var}\left(\theta_5\right) & = & 0.21\\
\mathrm{E}\left(\theta_6\right) & = & 0.31 & \mathrm{var}\left(\theta_6\right) & = & 0.18.
\end{array}
\label{eqn:cats_prior}
\end{equation}

The prior means in (\ref{eqn:cats_prior}) are given by the assumed fixed values of \cite{DrovandiPettitt}, with prior standard deviations given by the corresponding standard errors found by \cite{riley2003}, inferred from 95\% confidence intervals.

\begin{figure}
\begin{center}
\caption{Plots summarising results from the parasite model. Plots (a) and (b) show the mean nested Monte Carlo approximation (under auxiliary likelihood) to the expected SIG and LR utilities, respectively, against $n$ for designs found under the three different approaches.}
\label{fig:cats_perf}
\includegraphics[scale=0.75, viewport = 0 250 510 510, clip = true]{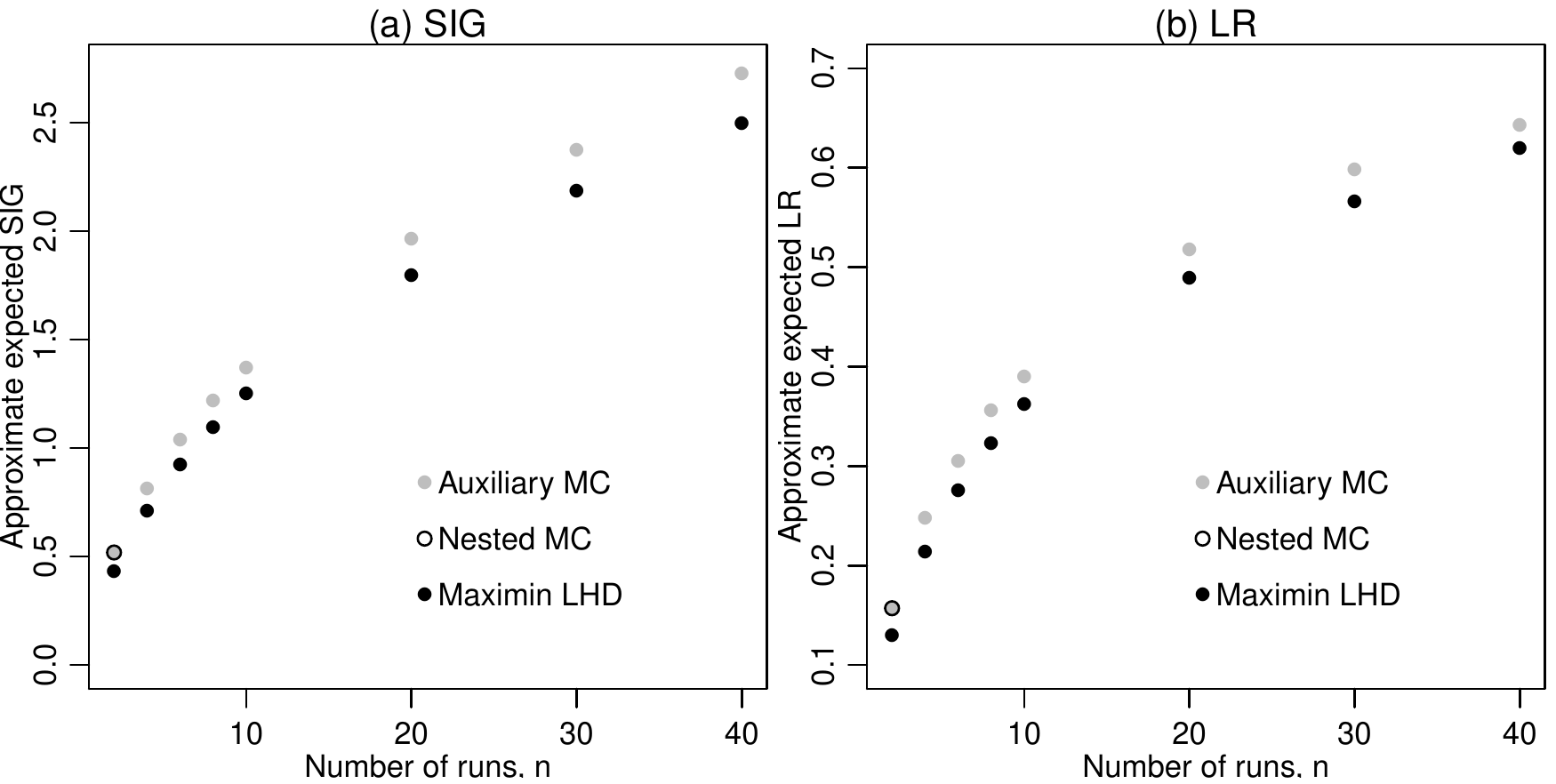}
\end{center}
\end{figure}

Since a mature parasite can only materialise from a juvenile, it means $y_k \in \left\{0, \dots, d_{k1} \right\}$. The obvious choice is to use a binomial distribution but similar to the aphid model in Section~\ref{sec:aphids}, this was under-dispersed compared to the parasite model. Instead, we choose $\mathcal{H}_{X}$ to be the beta-binomial distribution. This distribution was also used by \cite{cmryan2016}. The posterior predictive p-values for the conditional and marginal auxiliary models were $\mbox{p-value}_f = 0.20$ and $\mbox{p-value}_g = 0.11$, respectively. The posterior predictive p-values for the coupled auxiliary models ranged from 0.39 to 0.68 (over the different values of $n$ considered). Figure~\ref{fig:cats_check} in the Supplementary Material shows plots of sample statistics (mean and log variance) of the $\by^{(i)}_f$'s (the $\by^{(i)}_g$'s) against the $\by^{(i)}_{fX}$'s (the $\by^{(i)}_{gX}$'s). These plots and the posterior predictive p-values indicate that the auxiliary models are adequate.

We consider finding Bayesian designs under the SIG and LR utilities using ACE, under auxiliary Monte Carlo for each value of $n$. For $n=2$ we also use ACE to find a design using the nested Monte Carlo approximation (under the auxiliary likelihood). Similar to Section~\ref{sec:aphids}, we only consider $n=2$ due to the computational expense of finding designs under nested Monte Carlo for $n>2$. As a further comparison we also find maximin Latin hypercube designs (LHD) for each value of $n$. 

Figure~\ref{fig:cats_perf} shows the mean of twenty nested Monte Carlo approximations (under auxiliary likelihood) to the expected (a) SIG and (b) LR utilities against $n$ for the three different types of design. For both utilities, the auxiliary Monte Carlo designs are superior to the maximin Latin hypercube designs and, for $n=2$, there is negligible difference between the designs found under nested and auxiliary Monte Carlo. Table~\ref{tab:cats_times} in the Supplementary Material shows the average computing times required to find the designs under the SIG utility for auxiliary and nested Monte Carlo for each value of $n$. The computing time required to find a maximin Latin hypercube design is essentially negligible and not shown. Similar to Section~\ref{sec:aphids}, finding a design under nested Monte Carlo requires significantly more computing time than for auxiliary Monte Carlo.

\section{Model comparison} \label{sec:models}

\subsection{Bayesian design of experiments for model comparison} \label{sec:modelcomp}
Often interest lies in comparing a set $\mathcal{M}$ of competing stochastic models. An experimental aim of model comparison can be encapsulated by a utility function now denoted by $u(m,\mathbf{y},\mathbf{D})$ where $m \in \mathcal{M}$ denotes the unknown model. Fully Bayesian inference in this case centres on the posterior model probability of each model given by
$$\pi(m|\by,\bd) = \frac{ \pi(\by|\bd,m) \pi(m)}{\sum_{m \in \mathcal{M}}\pi(\by|\bd,m) \pi(m)},$$
where
\begin{equation}
\pi(\by|\bd,m) = \int_{\Theta_m} \pi(\mathbf{y}|\btheta_m,\mathbf{d},m) \pi(\btheta_m|m) \mathrm{d}\btheta_m
\label{eqn:margm}
\end{equation}
is the marginal likelihood and $\pi(m)$ the prior model probability, respectively, for model $m$. In (\ref{eqn:margm}), $\pi(\mathbf{y}|\btheta_m,\mathbf{d},m)$ is the likelihood for model $m$ with parameters $\btheta_m$ having prior distribution with pdf $\pi(\btheta_m|m)$. Two likelihood-based utility functions suitable for model comparison aims are
\begin{enumerate}
\item[(a)]
the SIG utility for models given by $u_{SM}(m,\mathbf{y},\mathbf{D}) = \log \pi(\mathbf{y}|m,\mathbf{D}) - \log \pi(\mathbf{y}|\mathbf{D})$, where $\pi(\mathbf{y}|\mathbf{D}) = \sum_{m \in \mathcal{M}} \pi(\mathbf{y}|m,\mathbf{D}) \pi(m)$; and
\item[(b)]
the 0-1 utility given by $u_{01}(m,\mathbf{y},\mathbf{D}) = I(m = \tilde{m})$ where $\tilde{m} = \arg \max_{m \in \mathcal{M}} \pi(m|\mathbf{y},\mathbf{D})$ is the posterior modal model.
\end{enumerate}

\subsection{Marginal auxiliary model} \label{sec:marge}

Both of the utility functions for model comparison given in Section~\ref{sec:modelcomp} depend on evaluation of the marginal likelihood, $\pi(\mathbf{y}|m,\mathbf{D})$, for each model $m \in \mathcal{M}$. Analogous to Section~\ref{sec:ml}, for $k=1,\dots,n$, let $\mathcal{G}(m,\mathbf{d})$ be the marginal distribution of $y_k$ for model $m$ having marginalised over the parameters $\btheta_m$. The marginal likelihood for model $m \in \mathcal{M}$ is then
$$\pi(\mathbf{y}|m,\bD) = c\left(G(y_1|m,\bd_1), \dots, G(y_n|m,\bd_n)|m,\bD\right) \times \prod_{k=1}^n g(y_k|m,\bd_k),$$
where $g(y_k|m,\bd_k)$ and $G(y_k|m,\bd_k)$ are the pdf/pmf and cdf of $\mathcal{G}(m,\mathbf{d})$, respectively, for $k=1,\dots,n$, and $c(\cdot|m,\bD)$ is the copula for the marginal model,  $\by|m,\bD$. 

A natural approach would be to find a separate coupled auxiliary model for each model, as described in Sections~\ref{sec:ml} and \ref{sec:copula}. However, one or more coupled auxiliary models may fit more adequately than the others, thus inflating the marginal likelihoods of these models. To mitigate this risk, we propose to find a separate copula for each model (as in Section~\ref{sec:meth}), but form a marginal auxiliary model which is dependent on $m$. Specifically, we assume $\mathcal{G}(m,\bd) = \mathcal{H}_X(\bphi_g(m,\bd))$ and set the marginal auxiliary model to be $\mathcal{G}_X(m,\bd) = \mathcal{H}_X(\hat{\bphi}_g(m,\bd))$ where $\hat{\bphi}_g(m,\bd)$ is an estimate of $\bphi_g(m,\bd)$ formed using an MGP. Full details are given in Section~\ref{app:newbit} in the Supplementary Material. The most important point is that the squared exponential correlation function given in (\ref{eqn:sqexp}) is only suitable for quantitative arguments, i.e. $\mathbf{d}$ or $\btheta$, but not $m$. \citet{qianetal2008} considered computer experiments where the arguments can be a mixture of quantitative and categorical. We adopt their exchangeable correlation function, i.e.
\begin{equation}
\kappa_{SEE}\left(\left(m^{(i)},\bd^{(i)}\right),\left(m^{(j)},\bd^{(j)}\right);\brho\right) = \exp \left( - \sum_{l=1}^w \rho_l \left( d_{il} - d_{jl}\right)^2 - \rho_{w+1} I(m^{(i)} \ne m^{(j)})\right).
\label{eqn:see}
\end{equation}

\subsection{Model comparison in epidemiological dynamics} \label{sec:epi}

We now consider a modified version of the model comparison example considered by \cite{dehideniya2017}. The set $\mathcal{M}$ refers to a set of different epidemiological models for the spread of a disease in a given population of known size $K=200$. The experiment involves observing $y_k$ the number of infected individuals in the population at time $t_k$, for $k=1,\dots,n$. Thus the design is $\bD = \left(t_1,\dots,t_n\right)$. The population also includes exposed and susceptible individuals. Exposed individuals are those who have been exposed to the disease but are not yet infected. Susceptible individuals are those who are at risk of becoming exposed. Let $S(t)$, $E(t)$, and $I(t)$ be the number of susceptible, exposed and infected individuals, respectively, at time $t$, constrained such that $S(t)+E(t)+I(t)=K$. Assume that at time $t=0$, $S(0)=K$ and $I(0)=E(0)=0$ and note that $y_k = I(t_k)$. \cite{dehideniya2017} considered the following four competing models.

\begin{enumerate}
\item
\textbf{Death model} ($m=1$)

In the death model, individuals transition from susceptible to infected directly, i.e. they do not become exposed as an intermediate step. The rate of transition is proportional to the number of susceptible individuals left in the population. These dynamics are given by
$$\mathrm{P}\left( S(t+\deltat) = S(t) - 1, I(t + \deltat) = I(t) +1 \right) = \theta_{11}S(t) \deltat + o(\deltat),$$
where $\btheta_1 = \left(\theta_{11}\right)$ is an unknown parameter.
\item
\textbf{Susceptible-Infected (SI) model} ($m=2$)

The SI model modifies the death model so that the rate of transition from susceptible to infected is proportional to the rate at which susceptible and infected individuals meet. These dynamics are given by
$$\mathrm{P}\left( S(t+\deltat) = S(t) - 1, I(t + \deltat) = I(t) +1 \right) = \left(\theta_{21} + \theta_{22} I(t)\right)S(t) \deltat + o(\deltat),$$
where $\btheta_2 = \left(\theta_{21},\theta_{22}\right)$ are unknown parameters.
\item
\textbf{Susceptible-Exposed-Infected (SEI) model} ($m=3$)

In the SEI model, the individuals can transition from susceptible to exposed to infected. The rate of these two transitions are proportional to the number of susceptible and exposed individuals, respectively. These dynamics are given by
\begin{eqnarray*}
& & \mathrm{P}\left( S(t+\deltat) = S(t) - 1, E(t+\deltat) = E(t)+1, I(t + \deltat) = I(t) \right) \\
& & \qquad \qquad \qquad \qquad  \qquad \qquad \qquad \qquad \qquad \qquad \qquad \qquad = \theta_{31} S(t) \deltat + o(\deltat),\\
& & \mathrm{P}\left( S(t+\deltat) = S(t) , E(t+\deltat) = E(t)-1, I(t + \deltat) = I(t)+1 \right) \\
& & \qquad \qquad \qquad \qquad  \qquad \qquad \qquad \qquad \qquad \qquad \qquad \qquad = \frac{E(t)}{\theta_{32}} \deltat + o(\deltat),
\end{eqnarray*}
where $\btheta_3 = \left(\theta_{31},\theta_{32}\right)$ are unknown parameters.
\item
\textbf{Susceptible-Exposed-Infected-II (SEI-II) model} ($m=4$)

The SEI-II model is a modification of the SEI model such that the rate of transition from susceptible to exposed is proportional to the rate at which susceptible and infected individuals meet. These dynamics are given by
\begin{eqnarray*}
& & \mathrm{P}\left( S(t+\deltat) = S(t) - 1, E(t+\deltat) = E(t)+1, I(t + \deltat) = I(t) \right) \\
& & \qquad \qquad \qquad \qquad  \qquad \qquad \qquad \qquad \qquad \qquad  = \left(\theta_{41} + \theta_{42}I(t)\right) S(t) \deltat + o(\deltat),\\
& & \mathrm{P}\left( S(t+\deltat) = S(t) , E(t+\deltat) = E(t)-1, I(t + \deltat) = I(t)+1 \right) \\
&&  \qquad \qquad \qquad \qquad  \qquad \qquad \qquad \qquad \qquad \qquad \qquad \qquad = \frac{E(t)}{\theta_{43}} \deltat + o(\deltat),
\end{eqnarray*}
where $\btheta_4 = \left(\theta_{41},\theta_{42},\theta_{43}\right)$ are unknown parameters.
\end{enumerate}

\begin{figure}
\begin{center}
\caption{Plots summarising results from the epidemiological dynamics example. Plots (a) and (b) show the mean nested Monte Carlo approximation (under auxiliary likelihood) to the expected SIG and 0-1 utilities, respectively, against $n$ for the design found under auxiliary Monte Carlo and the equally-spaced design. Plots (c) and (d) show the SIG and 0-1 designs, respectively, found under auxiliary Monte Carlo for each value of $n$.}
\label{fig:models_perf}
\includegraphics[scale=0.6]{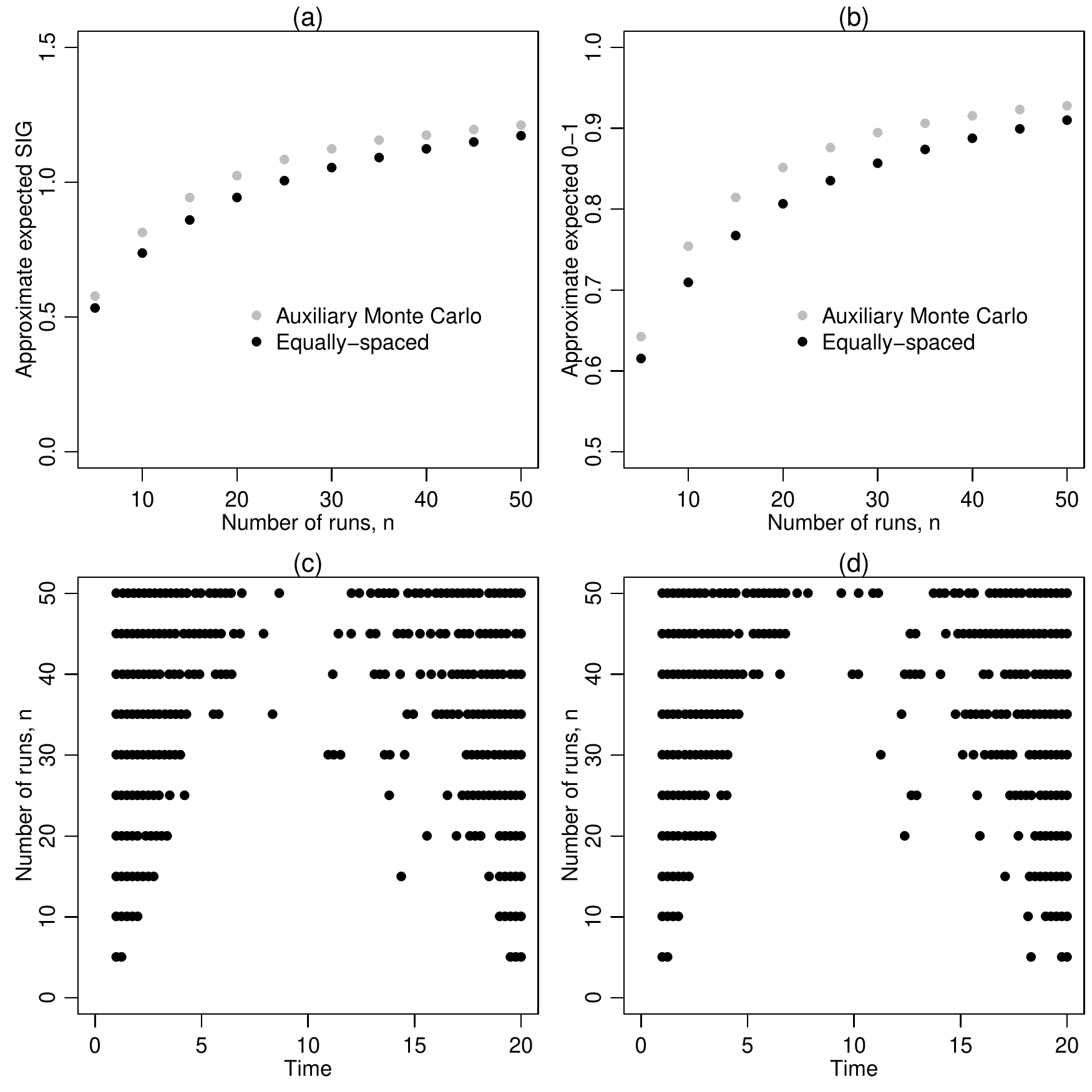}
\end{center}
\end{figure}

We consider finding designs under the two likelihood-based utilities described in Section~\ref{sec:modelcomp} for $n=5,10,\dots,50$.  \cite{dehideniya2017} also considered finding designs for the 0-1 utility function but used ABC to approximate the marginal likelihood and, therefore, only considered low-dimensional designs.

We set the prior model probabilities to be equal, i.e. $\pi(m) = 0.25$ therefore specifying that the models are \emph{a-priori} equally likely. For the prior distribution of $\btheta_m$ under each model $m$, we let each element of $\btheta_m$ have the following uniform distributions
\begin{equation}
\begin{array}{lcllcllcl}
\theta_{11} & \sim & \mathrm{U}[0,0.5], & & & & & & \\
\theta_{21} & \sim & \mathrm{U}[0,0.5], & \theta_{22} & \sim & \mathrm{U}[0,0.005], & & & \\
\theta_{31} & \sim & \mathrm{U}[0,0.5], & \theta_{32} & \sim & \mathrm{U}[0,10], & & & \\
\theta_{41} & \sim & \mathrm{U}[0,0.5], & \theta_{42} & \sim & \mathrm{U}[0,0.005], & \theta_{43} & \sim & \mathrm{U}[0,10].
\end{array}
\label{eqn:epi_priors}
\end{equation}
These were chosen so that the distribution of $\mathbf{y}|m,\mathbf{D}$ is approximately the same for all $m$. To see this, Figure~\ref{fig:models_prior} in the Supplementary Material shows samples of $\mathbf{y}$ plotted against time under each of the models. 

Since the number of infected individuals is bounded from above by $K$, it means that $y_k \in \left\{0,\dots, K\right\}$. Therefore, similar to the parasite model in Section~\ref{sec:cats}, we use the beta-binomial model for $\mathcal{H}_X$. The posterior predictive p-value for the marginal auxiliary model (for all $m$) is p-value$_g = 0.36$. The posterior predictive p-values for the coupled auxiliary models range from 0.20 to 0.41 over the different $n$ considered. Figure~\ref{fig:models_check} in the Supplementary Material shows plots of sample statistics (mean and variance) of the $\by^{(i)}_g$'s against the $\by^{(i)}_{gX}$'s. These plots appear to show some slight differences in the predictive variance between the four epidemiological dynamics models. However, in conjunction with the posterior predictive p-values, we consider the auxiliary models to be adequate.

The first row of Figure~\ref{fig:models_perf} shows the mean of twenty nested Monte Carlo approximations (under auxiliary likelihood) to the expected (a) SIG and (b) LR utilities against $n$ for the design found under auxiliary Monte Carlo and the design given by equally-spaced sampling times. For both utilities, the auxiliary Monte Carlo designs are superior to the equally-spaced designs. The second row of Figure~\ref{fig:models_perf} shows the (c) SIG and (d) 0-1 designs found under auxiliary Monte Carlo for each value of $n$. It can be seen in both cases, that the designs appear to have sampling times at the beginning and end of the sampling window.

\section{Discussion} \label{sec:disc}

In this paper we have introduced a general-purpose approach for finding Bayesian designs under intractable likelihood models. It is applicable for all likelihood-based utility functions, for realistic-sized experiments and for experimental aims of parameter estimation and model comparison. 

The proposed methodology is not applicable for non-likelihood-based utility functions. Examples of such utilities are the negative trace of the posterior variance matrix \citep[e.g.][]{overstall_woods_2016} or the determinant of the posterior precision matrix \citep[e.g][]{gillespieboys2018}. As stated by \citet{gillespieboys2018}, these utilities are only suitable for non-skewed unimodal posterior distributions. Likelihood-based utilities, on the other hand, use the likelihood function to characterise information coming from the experiment, as opposed to a single summary, such as the posterior mean or variance, and therefore make no restrictions on the posterior distribution.

\cite{cmryan2016} suggested the combination of auxiliary modelling and using normal-based approximations to posterior quantities \citep[e.g.][]{Laplace, overstall_normal}. However, as for our recommendation of likelihood-based utilities, we believe that the type of posterior distribution encountered in intractable likelihood models may not be well approximated by a normal distribution. Instead, other deterministic approximations, such as expectation propagation \citep[e.g.][pages 338-343]{gelman_etal_2014} could be more suitable.


\pagebreak

\begin{center}
\textbf{\large Supplementary Material for \textquotedblleft Bayesian design of experiments for intractable likelihood models using coupled auxiliary models and multivariate emulation\textquotedblright}
\end{center}

\setcounter{equation}{0}
\setcounter{figure}{0}
\setcounter{table}{0}
\setcounter{page}{1}
\makeatletter
\renewcommand{\theequation}{S\arabic{equation}}
\renewcommand{\thefigure}{S\arabic{figure}}
\renewcommand{\bibnumfmt}[1]{[S#1]}
\renewcommand{\citenumfont}[1]{S#1}

\section{Details on the approximate coordinate exchange algorithm} \label{app:ace}

\begin{enumerate}
\item
Choose an initial design $\mathbf{D}^0 = \left(D_1^0,\dots,D_q^0\right)$ and set the current design to be $\mathbf{D}^C = \left(D_1^C,\dots,D_q^C\right)=\mathbf{D}^0$. 
\item
For $i=1,\dots,q$ complete the following steps
\begin{enumerate}
\item
Let $U^i(D) = U(D^C_1,\dots,D^C_{i-1},D,D^C_{i+1},\dots,D^C_q)$ be the function given by the expected utility which only varies over the design space, $\mathcal{D}^i$, for the $i$th element.
\item \label{alg:approx} 
For $j=1,\dots,Q$, evaluate the Monte Carlo approximation to the expected utility given by
$$z_j = \hat{U}^i(D_j),$$
for $\left\{D_1,\dots,D_Q\right\} \in \mathcal{D}^i$. Fit a Gaussian process emulator to $\left\{z_j,d_j\right\}_{j=1}^Q$ and set $\tilde{U}^i(D)$ to be the resulting predictive mean.
\item
Find
$$D_i^* = \mathrm{arg} \mathrm{max}_{D^i \in \mathcal{D}} \tilde{U}^i(D),$$
and let $\mathbf{D}^* = \left(D^C_1,\dots,D^C_{i-1},D^*,D^C_{i+1},\dots\dots,D^C_q\right)$ be the proposed design.
\item \label{donaldtrump}
Set $\mathbf{D}^C = \mathbf{D}^*$ with probability $p^*$.
\end{enumerate}
\item
Return to step 2.
\end{enumerate}

In step \ref{donaldtrump}, we accept the proposed design, $\mathbf{D}^*$ with probability $p^*$. The proposed design originates from from the Gaussian process emulator. Similar to all statistical models, Gaussian process emulators can fit inadequately. To mitigate the effects of an inadequate emulator, \cite{overstall_woods_2016} proposed a comparison between the proposed design $\mathbf{D}^*$ and the current design $\mathbf{D}^C$ which is independent of the current Gaussian process emulator. Note that the proposed design $\mathbf{D}^*$ should be accepted if 
\begin{equation}
U (\mathbf{D}^*) > U (\mathbf{D}^C).
\label{eqn:compwert}
\end{equation}
For $b=1,\dots,B$ we generate samples $\left\{u^b_*\right\}_{b=1}^B$ and $\left\{u^b_C\right\}_{b=1}^B$ as follows
\begin{eqnarray*}
u^b_* & = & u (\mathbf{y}^{*b},\boldsymbol{\theta}^{*b},\mathbf{D}^*),\\
u^b_C & = & u (\mathbf{y}^b,\boldsymbol{\theta}^b,\mathbf{D}^C),
\end{eqnarray*}
where $\left\{\boldsymbol{\theta}^{*b},\mathbf{y}^{*b}\right\}_{b=1}^B$ and $\left\{\boldsymbol{\theta}^{b},\mathbf{y}^{b}\right\}_{b=1}^B$ are samples from the joint distribution of $\boldsymbol{\theta}$ and $\mathbf{y}$ conditional on $\mathbf{D}^*$ and $\mathbf{D}^C$, respectively. We use these samples to perform a Bayesian hypothesis test of (\ref{eqn:compwert}). The form of the Bayesian hypothesis test, as described in \cite{overstall_woods_2016}, assumes that the $u^b_*$'s and $u^b_C$'s are continuous and their distribution reasonably assumed normal. In this case, the probability of accepting the proposed design is
$$p^* = 1 - F\left(-\frac{B\bar{u}_* - B \bar{u}_C}{\sqrt{2B \hat{v}}}\right),$$
where $F(\cdot)$ is the distribution function of the $t$-distribution with $2B-2$ degrees of freedom,
$$\hat{v} = \frac{\sum_{b=1}^B (u_C^b - \bar{u}_C)^2 + \sum_{b=1}^B (u_*^b - \bar{u}_*)^2}{2B - 2},$$
and $\bar{u}_C$ and $\bar{u}_*$ are the sample means of the $u_C^b$'s and $u_*^b$'s, respectively.

The assumption of normality will clearly be violated for the 0-1 utility function for model comparison, described in Section~\ref{sec:modelcomp} of the main manuscript, where the $u^b_*$'s and $u^b_C$'s will be binary in the set $\left\{0,1\right\}$. For such utilities, 
\cite{overstall_normal} introduced a modification where 
$$p^* = 1 - \frac{1}{B} \sum_{b=1}^B F\left(\rho_C^b;1+B\bar{u}_*,1+B-B\bar{u}_*\right),$$
where $F\left(\cdot;a,b\right)$ denotes the distribution function of the $\mathrm{Beta}(a,b)$ and $\left\{\rho_C^b \right\}_{b=1}^B$ is a sample from \newline
$\mathrm{Beta}\left(1+B\bar{u}_C,1+B-B\bar{u}_C\right)$.

There are various controllable quantities (tuning parameters) in the ACE algorithm that need to be specified. We set $B=1000$ and $B=20000$ for fitting the GP model and the independent Bayesian hypothesis step, respectively. We also set the GP training sample to be $Q=20$. These are the default values in the \texttt{acebayes} \citep{acebayes} package. Additionally we found that twenty iterations of the ACE algorithm was sufficient to achieve approximate convergence in all examples. Finally, for each example, we restart the ACE algorithm from twenty different starting designs as proposed by \cite{overstall_woods_2016}. This is to mitigate against convergence to local optima.

\section{Construction of $\hat{\bphi}_g(\mathbf{d})$ for the marginal auxiliary model} \label{app:margaux}

The following description of how to construct $\hat{\bphi}_g(\mathbf{d})$ for the marginal auxiliary model closely follows the construction of $\hat{\bphi}_f$ for the conditional auxiliary model as described in Section~\ref{sec:mgp} of the Main Manuscript.

For $i=1,\dots,M$, we generate a sample, $\mathbf{y}_g^{(i)}$, of size $N$ from $\mathcal{G}(\mathbf{d}^{(i)})$ where $\left\{\mathbf{d}^{(i)}\right\}_{i=1}^M$ is the same training sample as described in Section~\ref{sec:mgp} of the Main Manuscript. For each of these samples, compute the MLE of $\bphi_g$ under $\mathcal{H}_X(\bphi_g)$, i.e.
$$\hat{\bphi}_g^{(i)} = \arg \max_{\bphi_g} \prod_{j=1}^N h_{X}(y_g^{(ij)}|\bphi_g).$$
Let $\mathbf{Z}_g$ be the $v \times M$ matrix with $i$th column given by $\bz_g^{(i)} = \lambda\left(\hat{\bphi}_g^{(i)}\right)$. Under the MGP, we assume
$$\mathbf{Z}_g|\bbeta_g, \bSigma_g, \mathbf{A}_g \sim \mathrm{MN}\left( \bbeta_g \mathbf{1}_{M}, \mathbf{A}_g, \bSigma_g \right),$$
where the $ij$th element of $\bA_g$ is given by (\ref{eqn:A}) (in the Main Manuscript) with $\mathbf{x}_i = \mathbf{d}^{(i)}$ and $\brho_f$ and $\eta_f$ replaced by $\brho_g$ and $\eta_g$, respectively, i.e. $s=w$. Now
$$\hat{\bphi}_g(\mathbf{d}) = \lambda^{-1} \left( \hat{\bbeta}_g + \left(\mathbf{Z}_g - \hat{\bbeta}_g \mathbf{1}_M\right) \hat{\bA}_g^{-1} \hat{\ba}_g\right),$$
where $\hat{\bbeta}_g$ is the MLE of $\bbeta_g$, $\hat{\bA}_g$ is $\bA_g$ with $\brho_g$ and $\eta_g$ replaced by their MLEs ($\hat{\brho}_g$ and $\hat{\eta}_g$, respectively), and $\hat{\ba}_g$ is an $M \times 1$ vector with $i$th element $\hat{a}_{gi} = \kappa(\mathbf{d}^{(i)}, \mathbf{d}; \hat{\brho}_g)$.

\section{Generating from the coupled auxiliary model} \label{app:couple}

To generate the sample $\left\{\breve{\mathbf{y}}_X^{(i)}\right\}_{i=1}^{M_0}$ from the coupled auxiliary model complete the following steps for $i=1,\dots,M_0$.  For $i=1,\dots,M_0$, these samples are generated under a design $\breve{\mathbf{D}}^{(i)} = \left(\breve{\mathbf{d}}^{(i1)},\dots,\breve{\mathbf{d}}^{(in)}\right)$, where each $\breve{\mathbf{d}}^{(ij)}$ is generated uniformly over $\mathcal{D}$.
\begin{enumerate}
\item
Generate $\breve{\btheta}^{(i)}$ from the prior distribution of $\btheta$.
\item
For $l=1,\dots,L$ and $k=1,\dots,n$, generate $y^{(lk)} \sim \mathcal{F}\left(\breve{\btheta}^{(i)}, \breve{\bd}^{(ik)}\right)$.
\item
Calculate
$$\hat{\bzeta}^{(i)} = \arg \max_{\bzeta} \prod_{l=1}^L c_{X}\left( G_{X}(\breve{y}^{(l1)}| \breve{\bd}^{(i1)}), \dots, G_{X}(\breve{y}^{(ln)}| \breve{\bd}^{(in)})|\bzeta\right).$$
\item
Generate $\bu^{(i)}$ from the copula, $\mathcal{C}_{X}$ with parameters $\hat{\bzeta}^{(i)}$. Now $\breve{y}_{X}^{(ik)} = G_{X}^{-1} \left( u^{(ik)}| \breve{\bd}^{(ik)}\right)$, where $G_{X}^{-1}$ is the inverse cdf of $\mathcal{G}_{X}$ and $u^{(ik)}$ is the $k$th element of $\mathbf{u}^{(i)}$. Set $\breve{\by}^{(i)}_{X} = \left(\breve{y}_{X}^{(i1)},\dots,\breve{y}_{X}^{(in)}\right)$.
\end{enumerate}

\section{Additional results from Section~\ref{sec:examples} of the Main Manuscript}

\begin{figure}
\begin{center}
\caption{Plots of sample mean of (a) the $\by^{(i)}_f$'s against the $\by^{(i)}_{fX}$'s and (b) the $\by^{(i)}_g$'s against the $\by^{(i)}_{gX}$'s, and plots of sample variances of (c) the $\by^{(i)}_f$'s against the $\by^{(i)}_{fX}$'s and (d) the $\by^{(i)}_g$'s against the $\by^{(i)}_{gX}$'s for the auxiliary models found for the compartmental model example in Section~\ref{sec:comp} of the Main Manuscript. In each plot a straight line of unit slope through the origin as been included as a reference.} \label{fig:comp_check}
\includegraphics[scale=0.9]{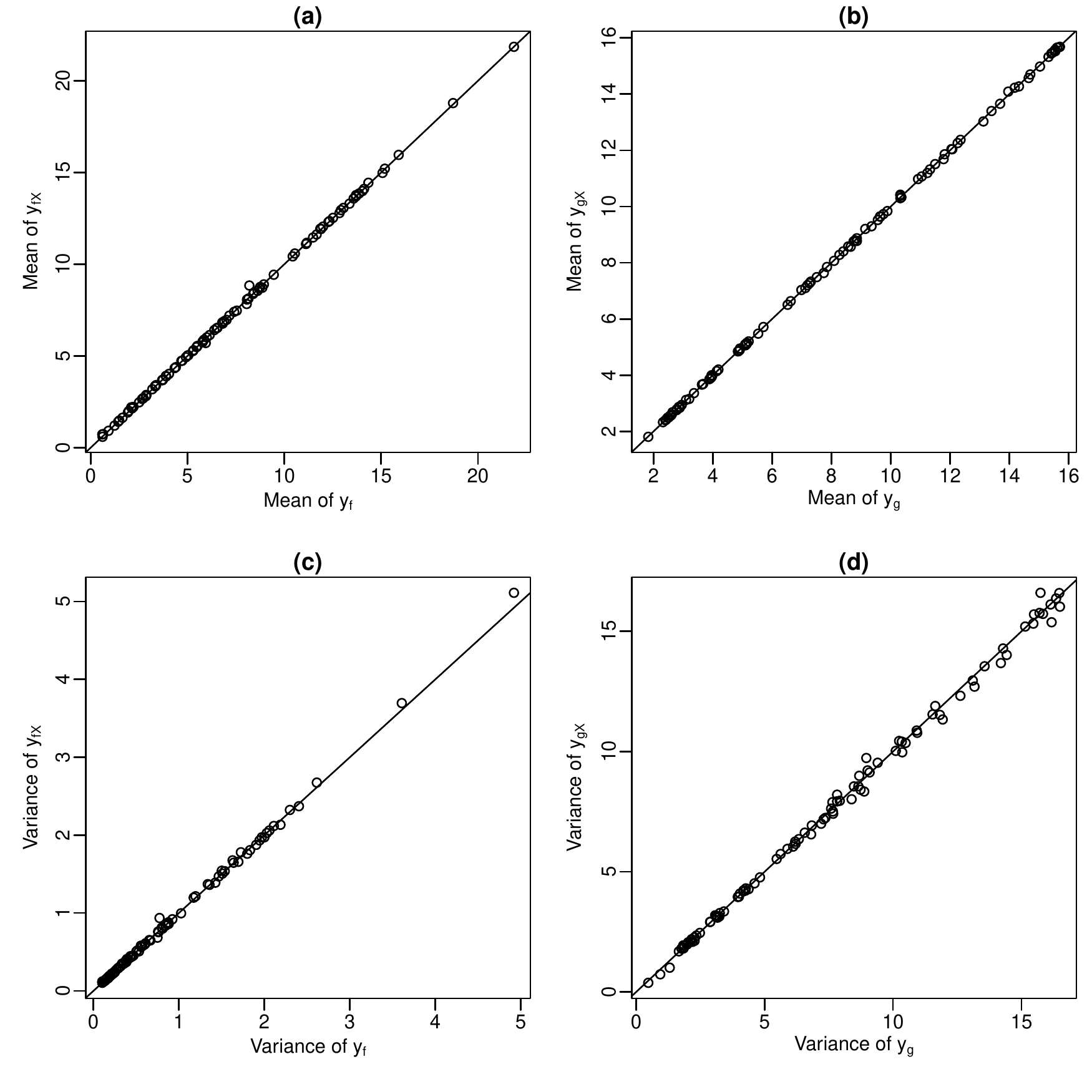}
\end{center}
\end{figure}

\begin{figure}
\begin{center}
\caption{Plots of sample mean of (a) the $\by^{(i)}_f$'s against the $\by^{(i)}_{fX}$'s and (b) the $\by^{(i)}_g$'s against the $\by^{(i)}_{gX}$'s, and plots of log sample variances of (c) the $\by^{(i)}_f$'s against the $\by^{(i)}_{fX}$'s and (d) the $\by^{(i)}_g$'s against the $\by^{(i)}_{gX}$'s for the beta binomial auxiliary models for the parasite model in Section~\ref{sec:cats} of the Main Manuscript. In each plot a straight line of slope one through the origin as been included as a reference.} \label{fig:cats_check}
\includegraphics[scale=0.9]{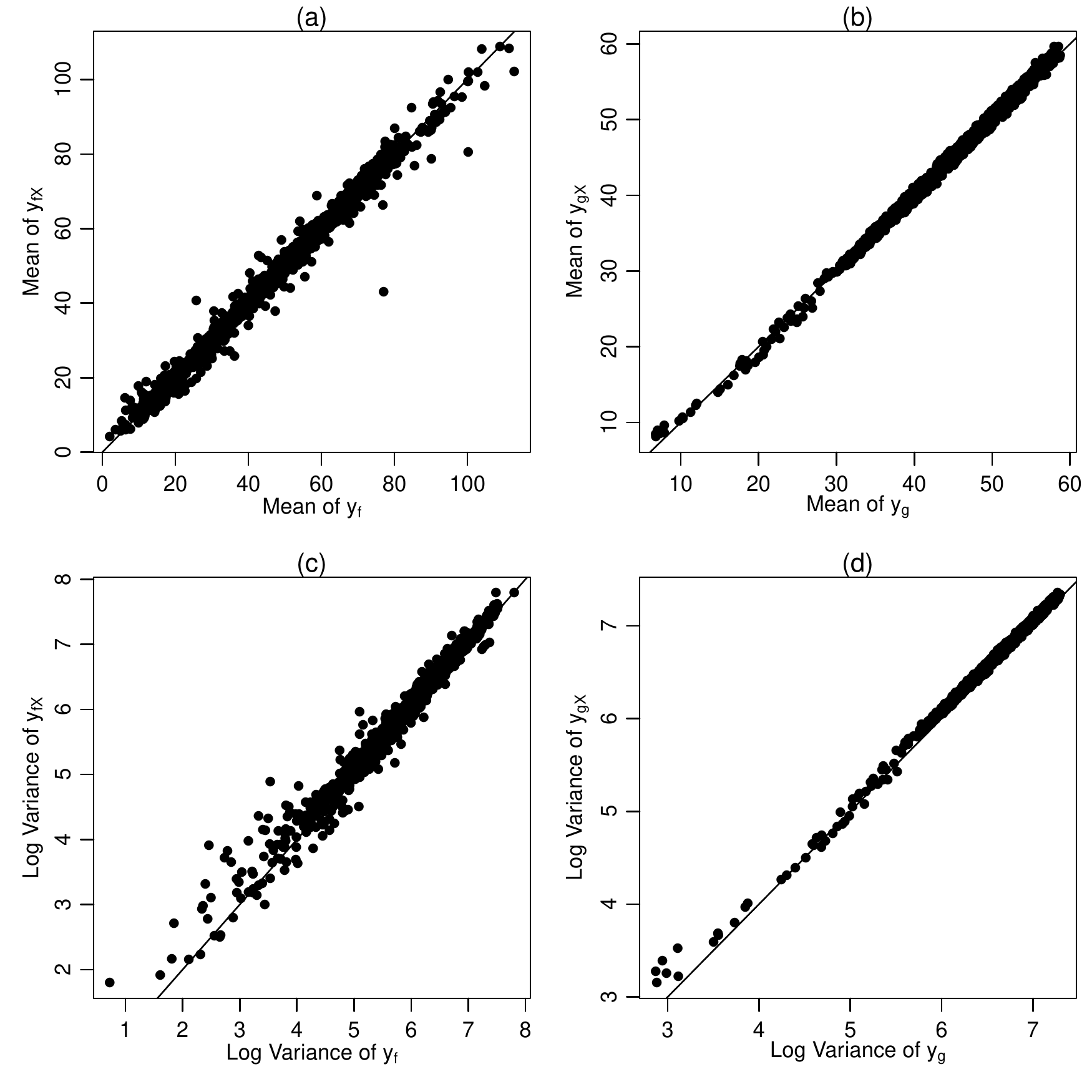}
\end{center}
\end{figure}

Figures~\ref{fig:comp_check} and~\ref{fig:cats_check} and Table~\ref{tab:cats_times} show additional results and are referred to in Section~\ref{sec:examples} of the Main Manuscript.

\begin{table}
\centering
\caption{Average computing time (in hours) for designs found under auxiliary and nested Monte Carlo for the SIG utility for the parasite model.} \label{tab:cats_times}
\begin{tabular}{lcccccccc}
Number of runs, $n$ & 2 & 4 & 6 & 8 & 10 & 20 & 30 & 40 \\ \hline
Auxiliary Monte Carlo & 0.2 & 0.8 & 1.6 & 2.8 & 4.2 & 16.1 & 34.9 & 57.8 \\
Nested Monte Carlo & 6.2 & - & - & - & - & - & - & - \\ \hline
\end{tabular}
\end{table} 

\section{Construction of $\hat{\bphi}_g(m,\mathbf{d})$ for the marginal auxiliary model} \label{app:newbit}

The following describes how to construct $\hat{\bphi}_g(m,\mathbf{d})$ for the marginal auxiliary model. 

For $i=1,\dots,M$, generate $m^{(i)}$ from the prior distribution of $m$ and then generate $\mathbf{y}_g^{(i)} = \left(y_g^{(i1)}, \dots, y_g^{(iN)}\right)$ where $y_g^{(ij)} \sim \mathcal{G}(m^{(i)},\mathbf{d}^{(i)})$ and $\left\{\mathbf{d}^{(i)}\right\}_{i=1}^M$ is the same training sample as described in Section~\ref{sec:mgp} of the Main Manuscript. For each of these samples, compute the MLE of $\bphi_g$ under $\mathcal{H}_X(\bphi_g)$, i.e.
$$\hat{\bphi}_g^{(i)} = \arg \max_{\bphi_g} \prod_{j=1}^N h_{X}(y_g^{(ij)}|\bphi_g).$$
Let $\mathbf{Z}_g$ be the $v \times M$ matrix with $i$th column given by $\bz_g^{(i)} = \lambda\left(\hat{\bphi}_g^{(i)}\right)$. Under the MGP, we assume
$$\mathbf{Z}_g|\bbeta_g, \bSigma_g, \mathbf{A}_g \sim \mathrm{MN}\left( \bbeta_g \mathbf{1}_{M}, \mathbf{A}_g, \bSigma_g \right),$$
where the $ij$th element of $\bA_g$ is given by 
$$A_{gij} = \kappa_{SEE}((m^{(i)},\mathbf{d}^{(i)}),(m^{(i)},\mathbf{d}^{(i)});\brho_g) + \eta_g I(i=j),$$
with $\kappa_{SEE}$ defined in (\ref{eqn:see}) (in the Main Manuscript). Now
$$\hat{\bphi}_g(m, \mathbf{d}) = \lambda^{-1} \left( \hat{\bbeta}_g + \left(\mathbf{Z}_g - \hat{\bbeta}_g \mathbf{1}_M\right) \hat{\bA}_g^{-1} \hat{\ba}_g\right),$$
where $\hat{\bbeta}_g$ is the MLE of $\bbeta_g$, $\hat{\bA}_g$ is $\bA_g$ with $\brho_g$ and $\eta_g$ replaced by their MLEs ($\hat{\brho}_g$ and $\hat{\eta}_g$, respectively), and $\hat{\ba}_g$ is an $M \times 1$ vector with $i$th element $\hat{a}_{gi} = \kappa_{SEE}(\left(m^{(i)},\mathbf{d}^{(i)}\right), \left(m,\mathbf{d}\right); \hat{\brho}_g)$.

\section{Additional results from Section~\ref{sec:models} of the Main Manuscript}

\begin{figure}
\begin{center}
\caption{Samples of $\mathbf{y}$ plotted against time under each of the models and the prior distribution given by (\ref{eqn:epi_priors}) (in the Main Manuscript)}
\label{fig:models_prior}
\includegraphics[scale=0.9]{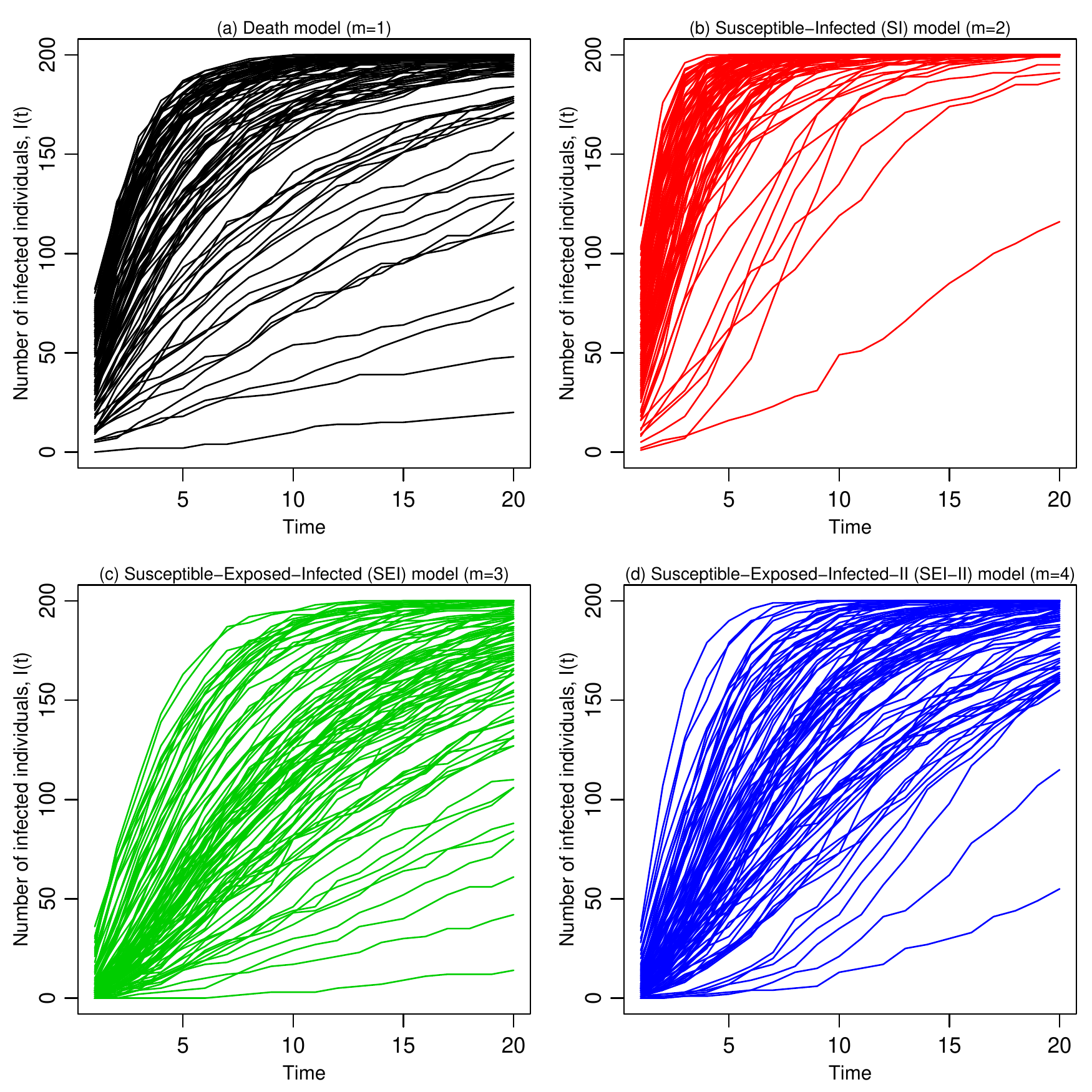}
\end{center}
\end{figure}

\begin{figure}
\begin{center}
\caption{Plots of (a) sample mean and (b) sample variance of the $\by^{(i)}_g$'s against the $\by^{(i)}_{gX}$'s for the marginal auxiliary model found for the epidemiological dynamics models in Section~\ref{sec:epi} of the Main Manuscript. The different colours indicate the model. In each plot a straight line of unit slope through the origin as been included as a reference.}
\label{fig:models_check}
\includegraphics[scale=0.9, viewport = 0 250 510 510, clip = true]{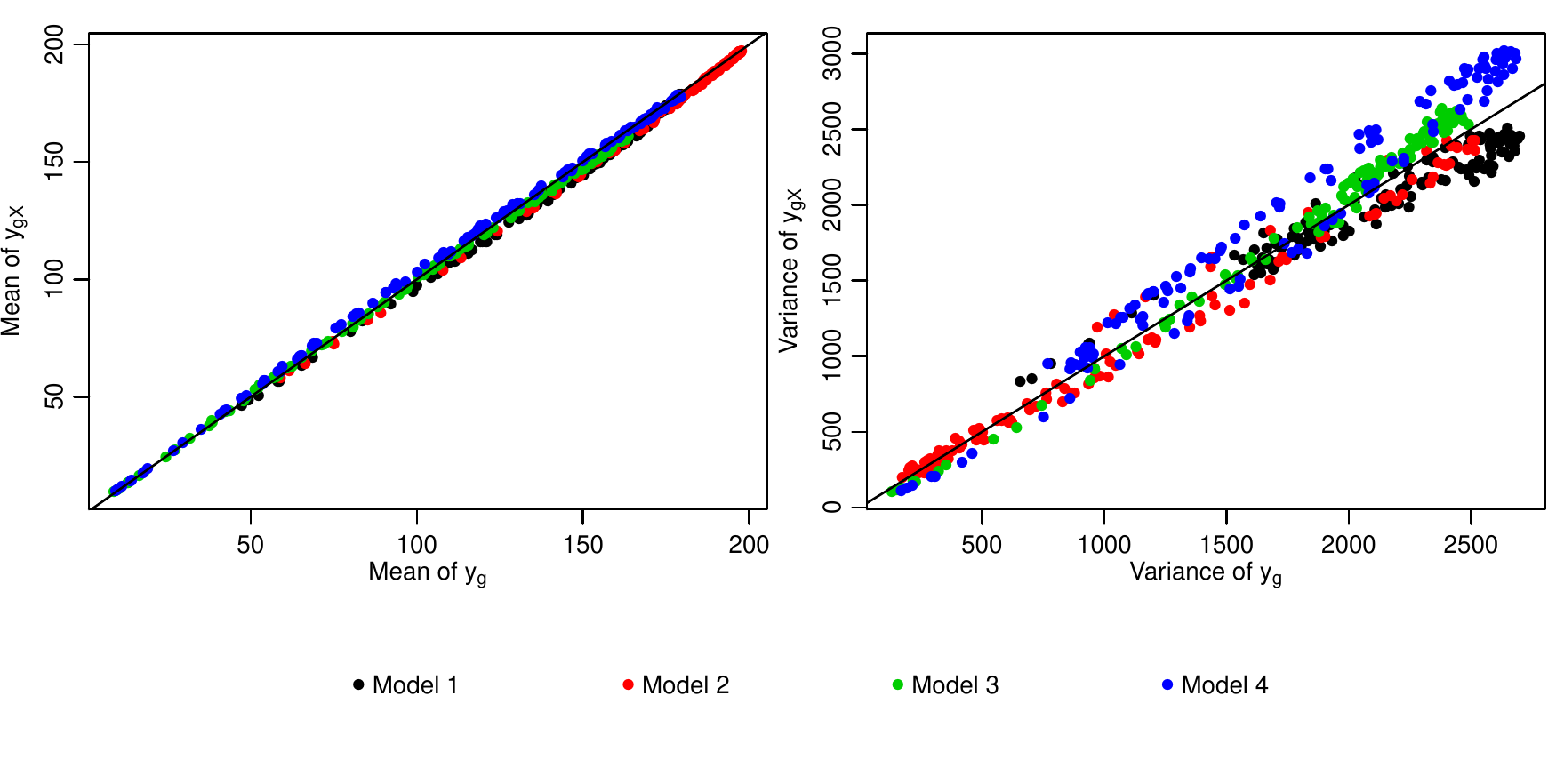}
\end{center}
\end{figure}

Figures~\ref{fig:models_prior} and~\ref{fig:models_check} show additional results and are referred to in Section~\ref{sec:epi} of the Main Manuscript.

\newpage

\bibliographystyle{rss}
\bibliography{biblio}

\begin{thebibliography}{50}
\expandafter\ifx\csname natexlab\endcsname\relax\def\natexlab#1{#1}\fi
\expandafter\ifx\csname url\endcsname\relax
  \def\url#1{\texttt{#1}}\fi
\expandafter\ifx\csname urlprefix\endcsname\relax\def\urlprefix{URL }\fi

\bibitem[{Abramowitz and Stegun(2002)}]{abra_ste}
Abramowitz, M. and Stegun, I.~A. (2002) Handbook of mathematical functions.
\newblock Available at \texttt{www.math.sfu.ca/$\sim$cbm/aands/toc.htm}.

\bibitem[{Amzal et~al.(2006)Amzal, Bois, Parent and
  Robert}]{amzal_bois_parent_robert_2006}
Amzal, B., Bois, F.~Y., Parent, E. and Robert, C.~P. (2006) Bayesian-optimal
  design via interacting particle systems.
\newblock \textit{Journal of the American Statistical Association}
  \textbf{101}, 773--785.

\bibitem[{Atkinson et~al.(2007)Atkinson, Donev and
  Tobias}]{atkinson_donev_tobias_2007}
Atkinson, A.~C., Donev, A.~N. and Tobias, R.~D. (2007) \textit{Optimum
  experimental designs, with {SAS}}.
\newblock Oxford University Press Inc., New York.

\bibitem[{Chaloner and Verdinelli(1995)}]{ChalonerVerdinelli}
Chaloner, K. and Verdinelli, I. (1995) Bayesian experimental design: A review.
\newblock \textit{Statistical Science} \textbf{10}, 273--304.

\bibitem[{Conti and O'Hagan(2010)}]{Conti2010}
Conti, S. and O'Hagan, A. (2010) Bayesian emulation of complex multi-output and
  dynamic computer models.
\newblock \textit{Journal of Statistical Planning and Inference} \textbf{140},
  640--651.

\bibitem[{Dean et~al.(2015)Dean, Morris, Stufken and Bingham}]{Dean2015}
Dean, A., Morris, M., Stufken, J. and Bingham, D. (eds.) (2015)
  \textit{Handbook of Design and Analysis of Experiments}.
\newblock Boca Raton: CRC Press.

\bibitem[{Dehideniya et~al.(2018)Dehideniya, Drovandi and
  McGree}]{dehideniya2017}
Dehideniya, M., Drovandi, C.~C. and McGree, J.~M. (2018) Optimal {B}ayesian
  design for discriminating between models with intractable likelihoods in
  epidemiology.
\newblock \textit{Computational Statistics and Data Analysis} \textbf{124},
  277--297.

\bibitem[{Demarta and McNeil(2005)}]{demartamcneil2005}
Demarta, S. and McNeil, A. (2005) The t copula and related copulas.
\newblock \textit{International Statistical Review} \textbf{73}, 111--129.

\bibitem[{Denham et~al.(1977)Denham, Ponnudurai, Nelson, Guy and
  Rogers}]{denham1977}
Denham, D., Ponnudurai, T., Nelson, G., Guy, F. and Rogers, R. (1977) Studies
  with brugia pahangi. i. parasitological observations on primary infections of
  cats (felis catus).
\newblock \textit{Journal for Parasitology} \textbf{2}, 239--247.

\bibitem[{Drovandi and Pettitt(2013)}]{DrovandiPettitt}
Drovandi, C.~C. and Pettitt, A.~N. (2013) Bayesian experimental design for
  models with intractable likelihoods.
\newblock \textit{Biometrics} \textbf{69(4)}, 937--948.

\bibitem[{Drovandi et~al.(2011)Drovandi, Pettitt and Faddy}]{Drovandi2011}
Drovandi, C.~C., Pettitt, A.~N. and Faddy, M.~J. (2011) Approximate {B}ayesian
  computation using indirect inference.
\newblock \textit{Journal of the Royal Statistical Society: Series C (Applied
  Statistics)} \textbf{60(3)}, 503--524.

\bibitem[{Drovandi et~al.(2015)Drovandi, Pettitt and Lee}]{drovandi2015}
Drovandi, C.~C., Pettitt, A.~N. and Lee, A. (2015) Bayesian indirect inference
  using a parametric auxiliary model.
\newblock \textit{Statistical Science} \textbf{30}, 72--95.

\bibitem[{Gelman et~al.(2014)Gelman, Carlin, Stern, Dunson, Vehtari and
  Rubin}]{gelman_etal_2014}
Gelman, A., Carlin, J., Stern, H., Dunson, D., Vehtari, A. and Rubin, D. (2014)
  \textit{Bayesian Data Analysis}.
\newblock Chapman and Hall 3rd edn.

\bibitem[{Gillespie and Boys(2019)}]{gillespieboys2018}
Gillespie, C. and Boys, R. (2019) Efficient construction of bayes optimal
  designs for stochastic process models.
\newblock \textit{Statistics and Computing} \textbf{In press}.

\bibitem[{Gillespie and Golightly(2010)}]{gillespiegolightly2010}
Gillespie, C. and Golightly, A. (2010) Bayesian inference for generalized
  stochastic population growth models with application to aphids.
\newblock \textit{Journal of Royal Statistical Society, Series C} \textbf{59},
  341--357.

\bibitem[{Gillespie(1977)}]{gillespie1977}
Gillespie, D. (1977) Exact stochastic inference of coupled chemical reactions.
\newblock \textit{Journal of Physical Chemistry} \textbf{81}.

\bibitem[{Gourieroux et~al.(1993)Gourieroux, Monfort and
  Renault}]{gourieroux1993}
Gourieroux, C., Monfort, A. and Renault, E. (1993) Indirect inference.
\newblock \textit{Journal of Applied Econometrics} \textbf{8}, 85--118.

\bibitem[{Hainy et~al.(2013)Hainy, M\"{u}ller and Wagner}]{hainy_etal_2013}
Hainy, M., M\"{u}ller, W. and Wagner, H. (2013) Likelihood-free
  simulation-based optimal design: An introduction.
\newblock In \textit{Topics in Statistical Simulation} (eds. V.~Melas,
  S.~Mignani, P.~Monari and L.~Salmaso) 271--278.

\bibitem[{Heggland and Frigessi(2004)}]{heggland2004}
Heggland, K. and Frigessi, A. (2004) Estimating functions in indirect
  inference.
\newblock \textit{Journal of the Royal Statistical Society Series B}
  \textbf{104}, 117--131.

\bibitem[{Huan and Marzouk(2013)}]{Huan2013}
Huan, X. and Marzouk, Y.~M. (2013) Simulation-based optimal {B}ayesian
  experimental design for nonlinear systems.
\newblock \textit{Journal of Computational Physics} \textbf{232(1)}, 288--317.

\bibitem[{Joe(1997)}]{joe1997}
Joe, H. (1997) \textit{Multivariate models and dependence concepts}.
\newblock Chapman \& Hall.

\bibitem[{Jones et~al.(2016)Jones, Goldstein, Jonathan and Randell}]{Jones}
Jones, M., Goldstein, M., Jonathan, P. and Randell, D. (2016) {B}ayes linear
  analysis for {B}ayesian optimal experimental design.
\newblock \textit{Journal of Statistical Planning and Inference} \textbf{171},
  115--129.

\bibitem[{Lange(2013)}]{lange_2013}
Lange, K. (2013) \textit{Optimization}.
\newblock Springer 2nd edn.

\bibitem[{Lindley(1956)}]{Lindley1956}
Lindley, D. (1956) On a measure of the information provided by an experiment.
\newblock \textit{Annals of Mathematical Statistics} \textbf{27}, 986--1005.

\bibitem[{Long et~al.(2013)Long, Scavino, Tempone and Wang}]{Laplace}
Long, Q., Scavino, M., Tempone, R. and Wang, S. (2013) Fast estimation of
  expected information gains for {B}ayesian experimental designs based on
  {L}aplace approximations.
\newblock \textit{Computer Methods in Applied Mechanics and Engineering}
  \textbf{259}, 24--39.

\bibitem[{Matis et~al.(2007)Matis, Kiffe, Matis and Stevenson}]{matis2007}
Matis, J., Kiffe, T., Matis, T. and Stevenson, D. (2007) Stochastic modeling of
  aphid population growth with nonlinear, power-law dynamics.
\newblock \textit{Mathematical Biosciences} \textbf{208}, 469--494.

\bibitem[{Meyer and Nachtsheim(1995)}]{MeyerNachtsheim}
Meyer, R. and Nachtsheim, C. (1995) The coordinate-exchange algorithm for
  constructing exact optimal experimental designs.
\newblock \textit{Technometerics} \textbf{37(1)}, 60--69.

\bibitem[{M\"{u}ller(1999)}]{Muller1999}
M\"{u}ller, P. (1999) Simulation-based optimal design.
\newblock \textit{Bayesian Statistics} \textbf{6}, 459--474.

\bibitem[{M\"{u}ller and Parmigiani(1995)}]{MullerParmigiani}
M\"{u}ller, P. and Parmigiani, G. (1995) Optimal design via curve fitting of
  {M}onte {C}arlo experiments.
\newblock \textit{Journal of the American Statistical Association}
  \textbf{90(432)}, 1322--1330.

\bibitem[{M\"uller et~al.(2004)M\"uller, Sanso and {De
  Iorio}}]{muller_sanso_iorio_2004}
M\"uller, P., Sanso, B. and {De Iorio}, M. (2004) Optimal {B}ayesian design by
  inhomogeneous {M}arkov chain simulation.
\newblock \textit{Journal of the American Statistical Association} \textbf{99},
  788--798.

\bibitem[{Nelson(1998)}]{nelson1998}
Nelson, R. (1998) \textit{An Introduction to Copulas}.
\newblock Springer Verlag New York.

\bibitem[{Overstall et~al.(2018{\natexlab{a}})Overstall, McGree and
  Drovandi}]{overstall_normal}
Overstall, A.~M., McGree, J.~M. and Drovandi, C.~C. (2018{\natexlab{a}}) An
  approach for finding fully {B}ayesian optimal designs using normal-based
  approximations to loss functions.
\newblock \textit{Statistics and Computing} \textbf{28}, 343--358.

\bibitem[{Overstall and Woods(2017)}]{overstall_woods_2016}
Overstall, A.~M. and Woods, D.~C. (2017) Bayesian design of experiments using
  approximate coordinate exchange.
\newblock \textit{Technometrics} \textbf{59}, 458--470.

\bibitem[{Overstall et~al.(2018{\natexlab{b}})Overstall, Woods and
  Adamou}]{acebayes}
Overstall, A.~M., Woods, D.~C. and Adamou, M. (2018{\natexlab{b}})
  \textit{acebayes: Optimal {B}ayesian Experimental Design using the {ACE}
  Algorithm}.
\newblock
  \urlprefix\url{https://cran.r-project.org/web/packages/acebayes/index.html}.
\newblock R package version 1.6.0.

\bibitem[{Pagendam and Pollett(2013)}]{pagendam2013}
Pagendam, D. and Pollett, P. (2013) Optimal design of experimental epidemics.
\newblock \textit{Journal of Statistical Planning and Inference} \textbf{143},
  563--572.

\bibitem[{Panagiotelis et~al.(2012)Panagiotelis, Czado and Joe}]{pana2012}
Panagiotelis, A., Czado, C. and Joe, H. (2012) Pair copula constructions for
  multivariate discrete data.
\newblock \textit{Journal of American Statistical Association} \textbf{107},
  1063--1072.

\bibitem[{Parker et~al.(2015)Parker, Gilmour, Schormans and
  Maruri-Aguilar}]{parker2015}
Parker, B., Gilmour, S., Schormans, J. and Maruri-Aguilar, H. (2015) Optimal
  design of measurements on queueing systems.
\newblock \textit{Queueing Systems} \textbf{79}, 365--390.

\bibitem[{Price et~al.(2016)Price, Bean, Ross and Tuke}]{price2016}
Price, D., Bean, N., Ross, J. and Tuke, J. (2016) On the efficient
  determination of optimal {B}ayesian experimental designs using {ABC}: A case
  study in optimal observation of epidemics.
\newblock \textit{Journal of Statistical Planning and Inference} \textbf{172},
  1--15.

\bibitem[{Qian et~al.(2008)Qian, Wu and Wu}]{qianetal2008}
Qian, P., Wu, H. and Wu, C. (2008) Gaussian process models for computer
  experiments with qualitative and quantitative factors.
\newblock \textit{Technometrics} \textbf{50}, 383--396.

\bibitem[{Riley et~al.(2003)Riley, Donnelly and Ferguson}]{riley2003}
Riley, S., Donnelly, C. and Ferguson, N. (2003) Robust parameter estimation
  techniques for stochastic within-host macroparasite models.
\newblock \textit{Journal of Theoretical Biology} 419--430.

\bibitem[{Ryan et~al.(2016{\natexlab{a}})Ryan, Drovandi and
  Pettitt}]{cmryan2016}
Ryan, C., Drovandi, C. and Pettitt, A.~N. (2016{\natexlab{a}}) Optimal
  {B}ayesian experimental design for models with intractable likelihoods using
  indirect inference applied to biological process models.
\newblock \textit{Bayesian Analysis} \textbf{11}, 857--883.

\bibitem[{Ryan et~al.(2016{\natexlab{b}})Ryan, Drovandi, McGree and
  Pettitt}]{ryan_drovandi_mcgree_pettitt_2015}
Ryan, E.~G., Drovandi, C.~C., McGree, J.~M. and Pettitt, A.~N.
  (2016{\natexlab{b}}) A review of modern computational algorithms for
  {B}ayesian optimal design.
\newblock \textit{International Statistical Review} \textbf{84}, 128--154.

\bibitem[{Ryan et~al.(2014)Ryan, Drovandi, Thompson and
  Pettitt}]{ryan_drovandi_thompson_pettitt_2014}
Ryan, E.~G., Drovandi, C.~C., Thompson, M.~H. and Pettitt, A.~N. (2014) Towards
  {B}ayesian experimental design for nonlinear models that require a large
  number of sampling times.
\newblock \textit{Computational Statistics and Data Analysis} \textbf{70},
  45--60.

\bibitem[{Ryan(2003)}]{Ryan2003}
Ryan, K. (2003) Estimating expected information gains for experimental designs
  with application to the random fatigue-limit model.
\newblock \textit{Journal of Computational and Graphical Statistics}
  \textbf{12}, 585--603.

\bibitem[{Santner et~al.(2003)Santner, Williams and Notz}]{santner_2003}
Santner, T., Williams, B. and Notz, W. (2003) \textit{The Design and Analysis
  of Computer Experiments}.
\newblock Springer, New York.

\bibitem[{Tavar\'{e} et~al.(1997)Tavar\'{e}, Balding, Griffiths and
  Donnelly}]{tavare1997}
Tavar\'{e}, S., Balding, D.~J., Griffiths, R. and Donnelly, P. (1997) Inferring
  coalescence times from {DNA} sequence data.
\newblock \textit{Genetics} \textbf{145}, 505--518.

\bibitem[{Tran et~al.(2017)Tran, Nott and Kohn}]{tran_nott_kohn_2015}
Tran, M., Nott, D.~J. and Kohn, R. (2017) Variational {B}ayes with intractable
  likelihood.
\newblock \textit{Journal of Computational and Graphical Statistics} To appear.

\bibitem[{Weaver et~al.(2016)Weaver, Williams, Anderson-Cook and
  Higdon}]{Weaver}
Weaver, B., Williams, B., Anderson-Cook, C. and Higdon, D. (2016) Computational
  enhancements to {B}ayesian design of experiments using {G}aussian processes.
\newblock \textit{Bayesian Analysis} \textbf{11}, 191--213.

\bibitem[{Wood(2010)}]{wood2010}
Wood, S. (2010) Statistical inference for noisy nonlinear ecological dynamic
  systems.
\newblock \textit{Nature} \textbf{466}, 1102--1104.

\bibitem[{Woods et~al.(2017)Woods, Overstall, Adamou and
  Waite}]{woods_ace_2016}
Woods, D., Overstall, A., Adamou, M. and Waite, T. (2017) Bayesian design of
  experiments for generalised linear models and dimensional analysis with
  industrial and scientific application.
\newblock \textit{Quality Engineering} \textbf{29}, 91--103.

\end{thebibliography}

\end{document}